\documentclass[twocolumn,showpacs,preprintnumbers, 
               superscriptaddress,nofootinbib]{revtex4-1}
\usepackage{graphicx}
\usepackage{amsfonts}
\usepackage{amssymb}
\usepackage{amsmath}
\usepackage{ulem}

\pdfoutput=1 

\begin{document} 

\title{Equation of State for SU(3) Gauge Theory \\
via the Energy-Momentum Tensor under Gradient Flow} 

\author{Masakiyo Kitazawa}
\email{kitazawa@phys.sci.osaka-u.ac.jp}
\affiliation{Department of Physics, Osaka University, 
  Toyonaka, Osaka 560-0043, Japan}
\affiliation{J-PARC Branch, KEK Theory Center,
  Institute of Particle and Nuclear Studies, KEK,
  203-1, Shirakata, Tokai, Ibaraki, 319-1106, Japan }

\author{Takumi Iritani}
\email{takumi.iritani@riken.jp}
\affiliation{Theoretical Research Division, Nishina Center, RIKEN, Wako 351-0198, Japan}

\author{Masayuki Asakawa}
\email{yuki@phys.sci.osaka-u.ac.jp}
\affiliation{Department of Physics, Osaka University, 
  Toyonaka, Osaka 560-0043, Japan}

\author{Tetsuo Hatsuda}
\email{thatsuda@riken.jp}
\affiliation{Theoretical Research Division, Nishina Center, RIKEN, Wako 351-0198, Japan}
\affiliation{Kavli IPMU (WPI), The University of Tokyo, Chiba 606-8502, Japan}

\author{Hiroshi Suzuki}
\email{hsuzuki@phys.kyushu-u.ac.jp}
\affiliation{Department of Physics, Kyushu University, 744 Motooka, Nishi-ku, Fukuoka, 819-0395, Japan}

\begin{abstract}
The energy density and the pressure of SU(3) gauge theory 
at finite  temperature are studied by direct lattice measurements of the 
renormalized energy-momentum tensor obtained by  the gradient flow. 
Numerical analyses are carried out with $\beta=6.287$--$7.500$ corresponding to the lattice
spacing $a= 0.013$--$0.061\,\mathrm{fm}$.
The spatial (temporal) sizes are chosen to be $N_s= 64$, $96$, $128$ ($N_{\tau}=12$, $16$, $20$, $22$, $24$)  with the aspect
ratio, $5.33 \le N_s/N_{\tau} \le 8$.
 Double extrapolation, $a\rightarrow 0$ (the continuum limit)  followed by  $t\rightarrow 0$
 (the zero flow-time limit), is taken  using the numerical data.
Above the critical  temperature, the thermodynamic
quantities are obtained with a few percent precision including statistical and  systematic errors.
The results are  in good agreement with previous high-precision data obtained by using
  the integral method.

\end{abstract}

\date{\today}
\preprint{RIKEN-QHP-236, KYUSHU-HET-170, J-PARC-TH-0071}

\maketitle
\flushbottom

\section{Introduction}

Thermodynamic observables in QCD such as the energy density $\varepsilon$  and the pressure 
$p$ as functions of temperature $T$ and baryon chemical potential $\mu_{\rm B}$ 
 provide fundamental information for studying the physics of 
relativistic heavy-ion collisions and compact stars.
 Because of their importance, high precision lattice simulations of $\varepsilon$  and $p$ in SU(3)  gauge theory 
\cite{Boyd:1996bx,Okamoto:1999hi,Umeda:2008bd,Borsanyi:2012ve,
Giusti:2015got} and in full QCD 
\cite{Borsanyi:2013bia,Bazavov:2014pvz} 
on the lattice at finite  $T$ have been carried out extensively for the past
few decades. In most of these studies   the integral method \cite{Boyd:1996bx}
 is adopted,  where     $\varepsilon$ and $p$ are obtained by 
integrating so-called the interaction measure $\Delta \equiv \varepsilon-3p$ calculated on the lattice.

Recently, a new method to calculate 
thermodynamic quantities  has been proposed  \cite{Suzuki:2013gza,Asakawa:2013laa}
on the basis of the gradient flow~\cite{Luscher:2010iy,Narayanan:2006rf,Luscher:2013vga,Borsanyi:2012zr,Fodor:2012td}.
In this method, one makes use of the renormalized energy-momentum tensor (EMT) operator
 $T_{\mu\nu}$ constructed from the ``flowed field''
at nonzero flow-time $t$ \cite{Suzuki:2013gza}.
Once EMT is defined,  $\varepsilon$  and $p$ can be calculated
 by simply taking thermal averages at any given temperature,
\begin{align}
\varepsilon = - \langle T_{44}\rangle , \quad 
p = \frac13 \sum_{i=1}^3 \langle T_{ii}\rangle .
\label{eq:e,p}
\end{align}
This method has been tested for the thermodynamics of 
SU(3) gauge theory in Ref.~\cite{Asakawa:2013laa} for the first time 
 with $\beta=6/g_0^2=5.89$--$6.56$ corresponding to the lattice
spacing $a= 0.041$--$0.11\,\mathrm{fm}$ and 
the  spatial (temporal) size $N_s= 32$ ($N_{\tau}=6$, $8$, $10$).
It was found that the $\varepsilon$ 
and $p$ obtained by the gradient flow with small statistics 
can be comparable  to those 
obtained by the integral method with high statistics. 
An extension of this method to full QCD has been also formulated
\cite{Makino:2014taa} and numerical results were 
reported recently \cite{Itou:2015gxx,Taniguchi:2016ofw}.

In the present paper, we report 
an improved analysis
of the 
thermodynamics of SU(3) gauge theory with the gradient flow.
Numerical analyses are carried out with $\beta=6.287$--$7.500$ corresponding to the lattice
spacing $a= 0.013$--$0.061\,\mathrm{fm}$.
The spatial (temporal) sizes are chosen to be  $N_s= 64$, $96$, $128$ ($N_{\tau}=12$, $16$, $20$, $22$, $24$)
 with the aspect ratio $5.33 \le N_s/N_{\tau} \le 8$.
 The double extrapolation, $a\rightarrow 0$ (the continuum limit)  followed by  $t\rightarrow 0$
 (the zero flow-time limit), is taken  using 
the data on these fine lattices.
We note that such a double 
 limit  could not be taken  in Ref.~\cite{Asakawa:2013laa} due to the coarse lattice.
 The lattice spacing $a$ required for these analyses has been  determined
on the basis of the gradient flow (see~Ref.~\cite{Asakawa:2015vta} and Appendix A.1).

After taking the double limit,  the final results of $\varepsilon$ and $p$ above the critical temperature~$T_c$
reach  a few percent precision with  both statistical error and systematic errors. The latter errors are 
associated with the $a\to 0$ and $t\to0$ extrapolations as well as  the  scale setting and lambda parameter. 
Our high precision results based on the gradient flow are found to be in  good agreement with the previous 
high precision results with the integral method.

This paper is organized as follows.
In the next section we introduce the gradient flow and 
the EMT operator used in our study.
After describing the setup of numerical simulations in Sec.~\ref{sec:setup},
the numerical results are presented in Sec.~\ref{sec:num}.
The last section is devoted to a short summary.
In Appendix~\ref{sec:scale}, the analyses of the lattice spacing and the 
lambda parameter are described in detail.

\section{Basic Formulation}
\label{sec:formalism}

\subsection{Yang--Mills gradient flow}
\label{sec:flow}

Let us first recapitulate  the essential features of the 
Yang--Mills gradient flow \cite{Luscher:2010iy} 
and its application to define the renormalized EMT \cite{Suzuki:2013gza}.

The gradient flow of  the Yang--Mills field is generated by the  differential equation 
 with a flow time  $t$, which has a dimension of inverse mass squared, 
\begin{align}
\frac{d A_\mu(t,x)}{dt} = - g_0^2 
\frac{ \delta S_{\rm YM}(t)}{ \delta A_\mu (t,x)}
= D_\nu G_{\nu\mu}(t,x) .
\label{eq:GF}
\end{align}
Here the Yang--Mills action $S_{\rm YM}(t)$ and 
the field strength $G_{\mu\nu}(t,x)$ are composed of the 
flowed field $A_\mu(t,x)$, 
which is a function of $t$ and the 4-dimensional Euclidean coordinate~$x$.
Color indices are suppressed for simplicity.
The initial condition at $t=0$ is taken to be $A_\mu(0,x)=A_\mu(x)$ with 
$A_\mu(x)$ being the ordinary gauge field in 4-dimensional Euclidean spacetime.

With Eq.~(\ref{eq:GF}), 
the gauge field flows along the steepest descent 
direction of $S_{\rm YM}(t)$ as $t$ increases.
At the tree level, Eq.~(\ref{eq:GF}) is rewritten as 
\begin{align}
  \frac{d A_\mu}{dt} = \partial_\nu \partial_\nu A_\mu
  + \text{(gauge dependent terms)},
\label{eq:diffusion}
\end{align}
which is a diffusion-type equation. Therefore, 
 the gradient flow for $t>0$ acts as 
a cooling of the gauge field with the smearing radius $\sqrt{8t}$ in the 
 4-dimensional Euclidean spacetime.
In Ref.~\cite{Luscher:2011bx}, it is  proved in pure gauge theory
that all  composite operators composed of~$A_\mu(t,x)$ take finite values for $t>0$.
Also, the idea of the gradient flow can be  generalized to gauge theory with fermions \cite{Luscher:2013cpa}.

\subsection{EMT from gradient-flow}
\label{sec:SFTE}

In the present study, we use the EMT defined by 
the short flow-time expansion \cite{Suzuki:2013gza}.
Let us consider a composite local operator ${O}(t,x)$ 
defined from the field $A_\mu(t,x)$ at positive flow time $t>0$.
The short flow-time expansion \cite{Luscher:2011bx} asserts 
that in the small $t$ 
limit this operator can be written as a superposition of 
local operators of the original gauge theory at $t=0$ as 
\begin{align}
  {O}(t,x) \xrightarrow[t\to0]{} \sum_i c_i(t) O_i^{\rm R}(x) ,
  \label{eq:SFTE}
\end{align}
where $O_i^{\rm R}(x)$ in the right-hand side are renormalized 
operators  of the original gauge 
theory at $t=0$ with the subscript $i$ denoting a set of operators,
while $c_i(t)$ are associated c-number coefficients calculable in 
perturbation theory for small $t$. 

In order to define the EMT using Eq.~(\ref{eq:SFTE}), 
we consider the short flow-time expansion of dimension-four 
gauge-invariant operators \cite{Suzuki:2013gza}.
In pure gauge theory, there are two such operators;
\begin{align}
  U_{\mu\nu}(t,x) &= G_{\mu\rho}^a (t,x)G_{\nu\rho}^a (t,x)
  -\frac14 \delta_{\mu\nu}G_{\rho\sigma}^a(t,x)G_{\rho\sigma}^a(t,x), 
  \label{eq:U}
  \\
  E(t,x) &= \frac14 G_{\mu\nu}^a(t,x)G_{\mu\nu}^a(t,x).
  \label{eq:E}
\end{align}
Since they are gauge invariant,
only gauge invariant operators appear in the 
right-hand side of  Eq.~(\ref{eq:SFTE}):
 Such an operator  with  dimension-zero  is an identity operator, while
 operators  with  dimension-four are EMTs $T_{\mu\nu}(x)$.
Up to this order, the short flow-time expansion
of Eqs.~(\ref{eq:U}) and (\ref{eq:E}) thus gives\footnote{This
useful combination was first given in~Ref.~\cite{DelDebbio:2013zaa}.}
\begin{align}
   U_{\mu\nu}(t,x)
   &=\alpha_U(t)\left[
   T_{\mu\nu}(x)-\frac14 \delta_{\mu\nu}T_{\rho\rho}(x)\right]
   +O(t),
\label{eq:UT}\\
   E(t,x)
   &=\left\langle E(t,x)\right\rangle_0
   +\alpha_E(t)T_{\rho\rho}(x)
   +O(t).
\label{eq:ET}
\end{align}
We normalize EMT so that the vacuum expectation values vanish, $\langle T_{\mu\nu}(x) \rangle_0 =0$.
This determines the coefficient of the unit operator in the right-hand side of Eq.~(\ref{eq:ET}) to be~$\langle E(t,x)\rangle_0$.  The unit operator does not appear in Eq.~(\ref{eq:UT}) since 
 $U_{\mu\nu}(t,x)$ is traceless.
 Contributions from the operators of  dimension six or higher are proportional to 
$t$ or higher from the dimensional reason, and thus they are suppressed for small $t$.

Combining relations Eqs.~(\ref{eq:UT}) and~(\ref{eq:ET}), we have
\begin{align}
  T_{\mu\nu}(x) =\lim_{t\to0} {T}_{\mu\nu}(t,x) ,
\label{eq:T^R}
\end{align}
with 
\begin{align}
  T_{\mu\nu}(t,x) = &
  \frac{1}{\alpha_U(t)}U_{\mu\nu}(t,x)
\nonumber \\
  &+\frac{\delta_{\mu\nu}}{4\alpha_E(t)}
  \left[E(t,x)-\left\langle E(t,x)\right\rangle_0 \right].
\label{eq:tildeT^R}
\end{align}
The coefficients $\alpha_U(t)$ and $\alpha_E(t)$ are calculated
perturbatively in the $\overline{\text{MS}}$ scheme 
in Ref.~\cite{Suzuki:2013gza},
\begin{align}
   \alpha_U(t)
   &=\Bar{g}(1/\sqrt{8t})^2
   \left[1+2b_0\Bar{s}_1\Bar{g}(1/\sqrt{8t})^2+O(\Bar{g}^4)\right],
\label{eq:a_U}
\\
   \alpha_E(t)
   &=\frac{1}{2b_0}\left[1+2b_0\Bar{s}_2
   \Bar{g}(1/\sqrt{8t})^2+O(\Bar{g}^4)\right],
\label{eq:a_E}
\end{align}
where $\Bar{g}(q)$ denotes the running gauge coupling in the
$\overline{\text{MS}}$ scheme with $q=1/\sqrt{8t}$ and
\begin{align}
  \bar{s}_1 &= \frac{7}{22} + \frac{1}{2}\gamma_E - \ln 2 \simeq
  - 0.08635752993, 
\\
  \bar{s}_2 &= \frac{21}{44} - \frac{b_1}{2b_0^2} = \frac{27}{484}
  \simeq
     0.05578512397,
\end{align}
with 
$b_0=\frac{1}{(4\pi)^2}\frac{11}{3}N_c$,
$b_1=\frac{1}{(4\pi)^4}\frac{34}{3}N_c^2$
with~$N_c=3$. 

We note here that 
(i)  the right-hand side of Eq.~(\ref{eq:T^R}) is 
independent of the regularization because of its UV finiteness,
so that one can take, e.g., the lattice regularization scheme,
and (ii) the small $t$ expansion of $U_{\mu \nu}(t,x)$ and $E(t,x)$ implies
\begin{align}
 {T}_{\mu\nu}(t,x) = T_{\mu\nu}(x) + {\cal O}(t).
  \label{eq:tildeTt}
\end{align}

\subsection{Energy density and pressure on the lattice}
\label{sec:ep}

The thermodynamic quantities
are obtained from the expectation values of diagonal elements of 
the EMT as in Eq.~(\ref{eq:e,p}).
A combination of $\varepsilon$ and~$p$ 
called the interaction measure~$\Delta$
 is related to the trace of the EMT (the trace
anomaly):
\begin{equation}
   \Delta=\varepsilon-3p
   =-\left\langle T_{\mu\mu}(x)\right\rangle.
\label{eq:Delta}
\end{equation}
Also, the entropy density~$s$ at zero chemical potential is 
given by $\varepsilon$ and $p$ as
\begin{equation}
   sT=\varepsilon+p
   =-\langle T_{44}(x)\rangle
   +\frac13 \sum_{i=1}^3 \langle T_{ii}(x)\rangle.
\label{eq:s}
\end{equation}

In the practical numerical analysis, we calculate 
Eq.~(\ref{eq:tildeT^R}) on a flowed gauge field with~$t>0$.
With finite $a$, the lattice gauge field has to be
smeared by the gradient flow sufficiently 
to suppress the lattice discretization effect.
Since the smearing length of the gradient flow is given by
$\sqrt{8t}$, this condition requires $\sqrt{8t}\gtrsim a$.
On the other hand, the value of $\sqrt{8t}$ has to be
small enough compared with half the temporal extent of 
the lattice, $1/(2T)$, so that the smearing by the gradient
flow does not feel the periodic boundary condition.
From these requirements, the measurement has to be
performed in the range
\begin{align}
a \lesssim \sqrt{8t} \lesssim \frac1{2T} .
\label{eq:t-range}
\end{align}
When Eq.~(\ref{eq:tildeT^R}) shows approximate  linear
dependence as shown in~Eq.~(\ref{eq:tildeTt})
in a range of $t$ given by Eq.~(\ref{eq:t-range}), 
one can take the small $t$ limit and obtain Eq.~(\ref{eq:T^R}).
The linear dependence Eq.~(\ref{eq:tildeTt}) can also be
violated for large $t$ when the perturbative results of 
the coefficients in Eqs.~(\ref{eq:a_U}) and (\ref{eq:a_E}) 
are no longer applicable.
This happens when $\sqrt{8t}$ approaches the lambda parameter 
$\Lambda_{\overline{\mathrm{MS}}}$.

For the measurement of $\Delta$, we have to calculate
$\langle E(t,x) \rangle_0$ to carry out vacuum subtraction.
This means that the numerical analysis for vacuum configuration
is needed in addition to $T>0$ simulation.
On the other hand, the analysis of $sT$, which depends only
on the traceless part, does not require
the vacuum subtraction and hence can be performed 
solely with a nonzero $T$ simulation.
This is an advantage of our method compared with the integral
method.\footnote{An alternative method to analyze $sT$ without vacuum 
subtraction is recently proposed in 
Ref.~\cite{Giusti:2012yj,Giusti:2015got}.}

\section{Simulation setup}
\label{sec:setup}

\begin{table}
\begin{tabular}{ccrrrc}
\hline \hline
$T/T_c$ & $\beta$ & $N_s$ & $N_\tau$ & confs. & vacuum \\
\hline
0.93 & 6.287 &  64 & 12 &  2125 & * \\
     & 6.495 &  96 & 16 &  1645 & * \\
     & 6.800 & 128 & 24 &  2040 & * \\
\hline
1.02 & 6.349 &  64 & 12 &  2000 & * \\
     & 6.559 &  96 & 16 &  1600 & * \\
     & 6.800 & 128 & 22 &  2290 & * \\
\hline
1.12 & 6.418 &  64 & 12 &  1875 & * \\
     & 6.631 &  96 & 16 &  1580 & * \\
     & 6.800 & 128 & 20 &  2000 & * \\
\hline
1.40& 6.582 &  64 & 12 &  2080 & * \\
     & 6.800 & 128 & 16 &   900 & * \\
     & 7.117 & 128 & 24 &  2000 & * \\
\hline
1.68 & 6.719 &  64 & 12 &  2000 & * \\
     & 6.941 &  96 & 16 &  1680 & * \\
     & 7.117 & 128 & 20 &  2000 & * \\
\hline
2.10 & 6.891 &  64 & 12 &  2250 & \\
     & 7.117 & 128 & 16 &   840 & * \\
     & 7.296 & 128 & 20 &  2040 & \\
\hline
2.31 & 7.200 &  96 & 16 &  1490 & \\
     & 7.376 & 128 & 20 &  2020 & \\
     & 7.519 & 128 & 24 &  1970 & \\
\hline
2.69 & 7.086 &  64 & 12 &  2000 & \\
     & 7.317 &  96 & 16 &  1560 & \\
     & 7.500 & 128 & 20 &  2040 & \\
\hline \hline
\end{tabular}

\caption{
Simulation parameters $\beta=6/g_0^2$, $N_s^3\times N_\tau$
and the number of configurations 
for nonzero temperature simulations at $T/T_c$.
The * symbol in the far right column shows the set of 
configurations that the corresponding vacuum simulation ($N_s=N_\tau$)
is available.
}
\label{table:param1}
\end{table}

\begin{table}
\begin{tabular}{crr}
\hline \hline
$\beta$ & $N_{s,\tau}$ & confs. \\
\hline
6.287 & 64 &  2125 \\
6.349 & 64 &   950 \\
6.418 & 64 &  1000 \\
6.582 & 64 &   800 \\
6.719 & 64 &  1000 \\
6.495 & 96 &   840 \\
6.559 & 96 &   840 \\
6.631 & 96 &   900 \\
6.941 & 96 &   837 \\
6.800 &128 &   992 \\
7.117 &128 &  1028 \\
\hline\hline
\end{tabular}

\caption{
Parameters for vacuum simulations ($N_s=N_\tau$).
}
\label{table:param2}
\end{table}

We have performed numerical simulations of SU(3) 
gauge theory on four-dimensional Euclidean lattices.
We considered the Wilson plaquette gauge action under the periodic
boundary condition with several different values of 
$\beta=6/g_0^2$ with $g_0$ being the bare coupling constant.
Gauge configurations are generated by the pseudo heat bath algorithm 
with the over-relaxation, mixed in the ratio of~$1:5$. 
We call one pseudo heat bath update plus five over-relaxation 
sweeps as a ``Sweep''. 
Each measurement is separated by $200$ Sweeps.
Statistical errors are then estimated by the jackknife method.
The binsize $N_{\rm bin}$ of the jackknife analysis is determined
so that the total number of jackknife bins is $50$ 
unless otherwise stated.
We have checked that the $N_{\rm bin}$ dependence of the 
statistical error is not observed with this binsize.

We use the Wilson gauge action for $S_{\rm YM}(t)$
 in  the  flow equation~Eq.~(\ref{eq:GF}).
The gradient flow in the $t$-direction is numerically solved by 
the third order Runge--Kutta (RK) method \cite{Luscher:2010iy}.
The RK time-step  is taken to be $0.01$ for small $t$ 
and is increased gradually as $t$ increases.
 Accumulation errors due to the RK
method is found to be more than two orders of magnitude  smaller than the statistical errors
in all the analyses discussed below.

For the operator $U_{\mu\nu}(t,x)$ in~Eq.~(\ref{eq:U}) 
necessary to analyze  $s/T^3$, we use 
$G^a_{\mu\nu}(t,x)$ written in terms of the clover leaf 
representation. 
For $E(t,x)$ in~Eq.~(\ref{eq:E}) 
necessary to analyze  $\Delta/T^4$,  
we use the   mixed representation~\cite{Fodor:2014cpa,Kamata:2016any},
 \begin{align}
E(t,x)_{\rm imp} = \frac34 E(t,x)_{\rm clover} + \frac14 E(t,x)_{\rm plaq},
\label{eq:E_imp}
\end{align}
where $E(t,x)_{\rm clover}$ is constructed from the clover leaf 
representation of $G^a_{\mu\nu}(t,x)$ in~Eq.~(\ref{eq:E}), while 
$E(t,x)_{\rm plaq} $ is defined as 
\cite{Luscher:2010iy}
\begin{align}
E(t,x)_{\rm plaq} = \frac1{18} P(t,x),
\label{eq:E_plaq}
\end{align}
with the plaquette $P(t,x) = 1/(6N_c) \sum_{\mu,\nu} {\rm Re \ Tr}$
$
[ U_{\mu}(t,x) U_{\nu}(t,x+\hat{\mu})
   U^{\dagger}_{\mu}(t,x+\hat{\nu}) U^{\dagger}_{\nu}(t,x) ]$.
 If the Wilson gauge action is employed for both the gauge action and 
$S_{\rm YM}(t)$ in Eq.~(\ref{eq:GF}), the $O(a^2)$ discretization errors in
$E(t,x)_{\rm imp} $ are cancelled  out in the tree level~\cite{Fodor:2014cpa}.

To specify temperature of a lattice as well as to perform 
the continuum extrapolation,
we need to relate  $\beta$ to the lattice spacing~$a$.
For this purpose, we  have previously performed measurements of
$a$  in the range 
$6.3\le\beta\le7.5$ using the gradient flow \cite{Asakawa:2015vta}.
As summarized in Appendix~\ref{sec:scale}, 
we derived a relation between the dimensionless reference scale $w_0/a$ and
 $\beta$ as
\begin{equation}
  \frac{w_0}a = \exp
  \left( \frac{4\pi^2}{33} \beta - 9.1268 + \frac{41.806}{\beta}
  - \frac{158.26}{\beta^2}\right) ,
  \label{eq:flowscale}
\end{equation}
which is applicable in the range $6.3 \leq \beta \leq 7.4$.
The statistical error of Eq.~(\ref{eq:flowscale}) associated with 
the fitting paramters is less than $0.4\%$.
Topological freezing of the data may also introduce extra
  $1\%$ error to this result (see Appendix~\ref{sec:scale1} for more details). 
To determine $T$ of a lattice in the unit of $T_c$,
we use the critical coupling after the infinite volume extrapolation
$\beta_c = 6.33552(47)$ at $N_\tau = 12$ 
\cite{Shirogane:2016zbf} and Eq.~(\ref{eq:flowscale}), which give
\begin{equation}
  w_0 T_c = 0.25244(17) .
  \label{eq:w04Tc}
\end{equation}

\begin{figure*}
  \includegraphics[width=0.45\textwidth,clip]{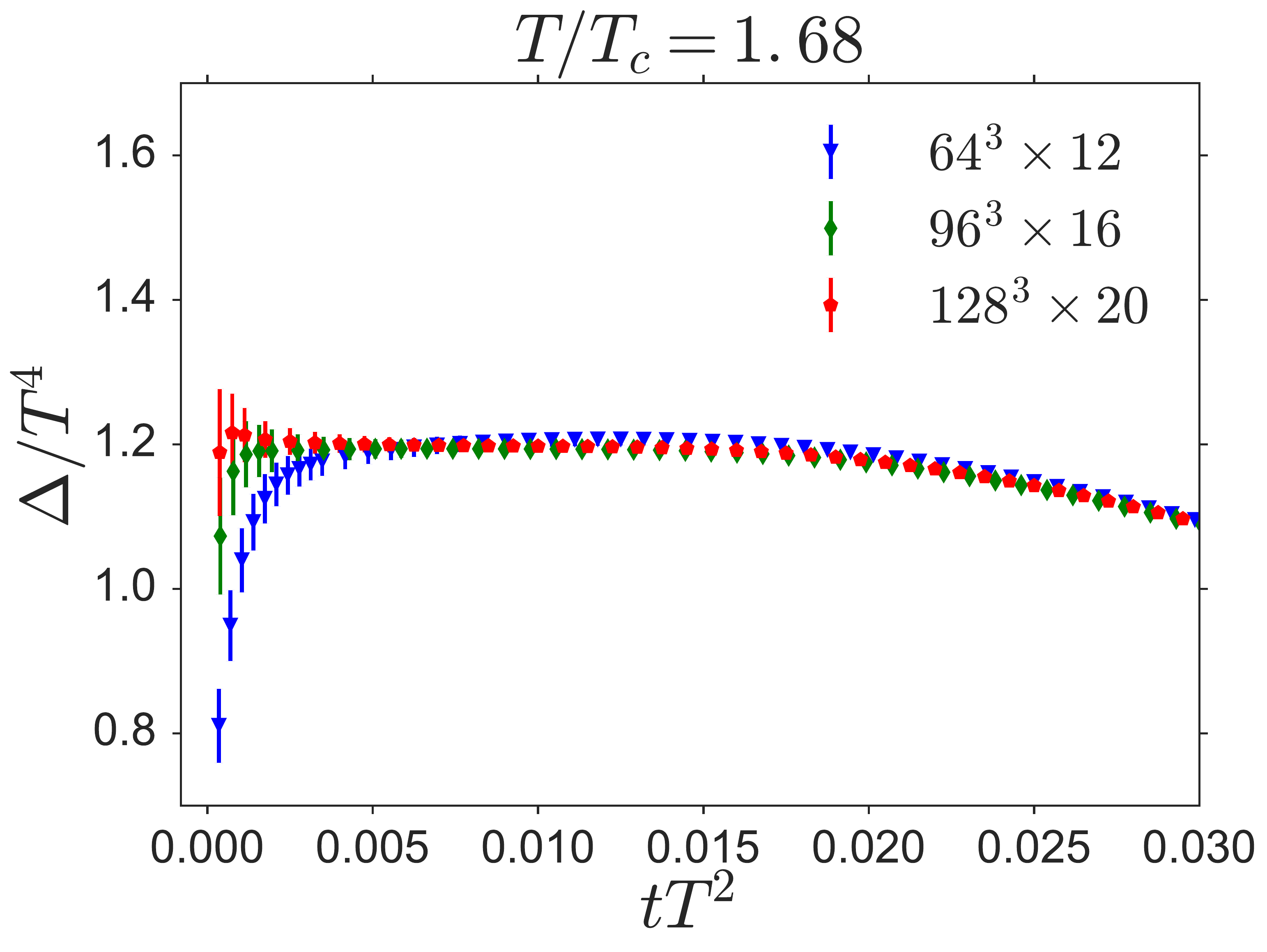}
  \includegraphics[width=0.45\textwidth,clip]{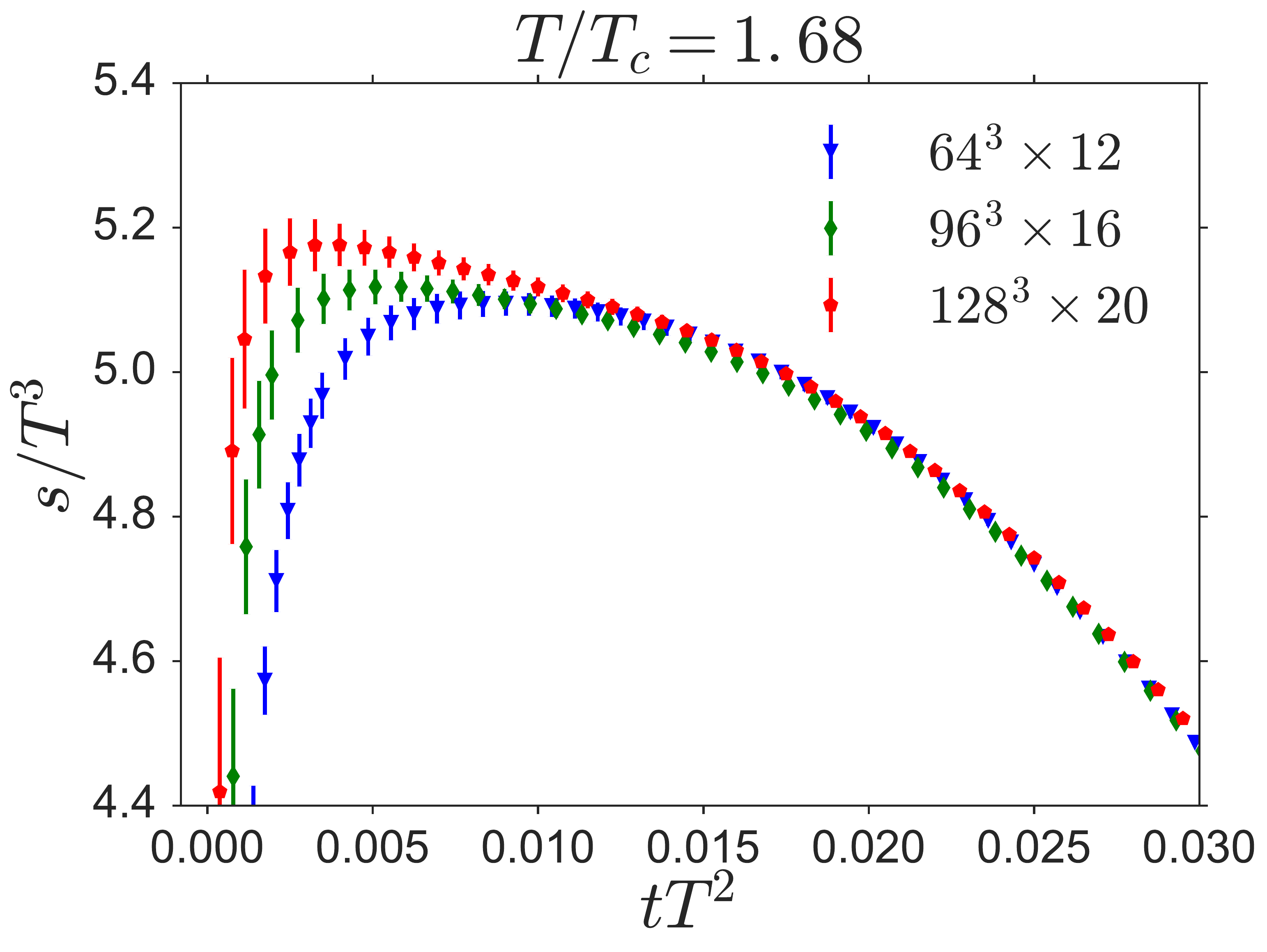}
\caption{
  Flow time $t$ dependences of trace anomaly $\Delta/T^4=(\varepsilon-3p)/T^4$
  (left) and entropy density $s/T^3=(\varepsilon+p)/T^4$ (right) for 
  $T/T_c=1.68$ with $N_\tau=12$, $16$ and~$20$.
}
\label{fig:166}
\end{figure*}

In the definition of the EMT operator Eq.~(\ref{eq:tildeT^R}), 
we need the running coupling $\bar{g}(q)$ in the 
$\overline{\mathrm{MS}}$ scheme which appears in the coefficients
$\alpha_U(t)$ and $\alpha_E(t)$ given by Eqs.~(\ref{eq:a_U}) 
and~(\ref{eq:a_E}).
To obtain $\bar{g}(q)$ at $q=1/\sqrt{8t}$, we need 
a functional form of $\bar{g}(q)$ and the relation between
the lattice spacing and $\Lambda_{\overline{\mathrm{MS}}}$.
We use the iterative formula for four-loop running coupling 
\cite{Agashe:2014kda} and 
\begin{equation}
  w_0\Lambda_{\overline{\mathrm{MS}}} = 0.2154(5)(11).
  \label{eq:w04Lambda}
\end{equation}
See Appendix \ref{sec:Lambda} as well as  Ref.~\cite{Asakawa:2015vta}
for more details.  Note that topological freezing would
introduce extra $1\%$ error to this result, too.

The simulation parameters are summarized in Tables~\ref{table:param1}
and \ref{table:param2}.
We perform the numerical simulations for eight different temperatures
in the range $0.93 \le T/T_c \le 2.69$ on the lattice of volume 
$N_s^3\times N_\tau$ as summarized in Table~\ref{table:param1}.
For each $T/T_c$, we perform numerical simulations
for three different values of $N_\tau$.
The value of $\beta$, lattice volume $N_s^3\times N_\tau$ and 
the number of configurations are shown in the table.
The aspect ratios $N_s/N_\tau$ of all lattices are 
within the range $5.33\le N_s/N_\tau \le 8$.
The values of $N_\tau$ for two coarse lattices are fixed to 
$N_\tau=12$ and $16$.
The finest lattice has the value of $N_\tau$ in the range 
$N_\tau=20$--$24$; because the corresponding vacuum simulation 
on $128^4$ lattice requires a large numerical cost, 
we make use of a single vacuum simulation for several values of $T$ 
by changing~$N_\tau$.

Since the lattice spacing determined by Eq.~(\ref{eq:flowscale})
has $1\%$ error, the value of $T/T_c$ on each lattice is 
expected to have a similar-size uncertainty.  Also, there is  a possible
finite volume effect, although  it is expected to be small
due to our large aspect ratio, $5.33 \le  N_s/N_\tau$.
These small uncertainties are not considered in the final results of
 $\Delta/T^4$ and $s/T^3$ to be shown at the end of this paper.

For the measurement of $\Delta/T^4$, we need the vacuum simulation
for vacuum subtraction.
We carry out the simulations on $N_\tau=N_s$ 
lattices corresponding to the temperatures 
in the range $0.93 \le T/T_c \le 1.68$.
The simulation parameters are shown in Table~\ref{table:param2}.
The configuration sets whose vacuum subtraction is available 
are shown by $*$ symbol in the far right column 
in Table~\ref{table:param1}.

To obtain the expectation values of the EMT with Eqs.~(\ref{eq:T^R})
and (\ref{eq:tildeT^R}), the double extrapolation $(t,a)\to(0,0)$
has to be taken.
To proceed this analysis, we first take the continuum limit,
$a\to0$,  with fixed $t$ in physical unit.
Since the leading lattice discretization effect on the thermodynamic 
quantities with the Wilson plaquette gauge action
 is of order $a^2$ \cite{Boyd:1996bx}, we take the following  parametrization to
 take  the continuum limit:
\begin{align} 
   \langle {T}_{\mu\nu}(t,x) \rangle_{{\rm lat}}
 =  \langle {T}_{\mu\nu}(t,x) \rangle_{\rm cont}
   + \frac{b_{\mu \nu}(t)}{N_\tau^2}.
  \label{eq:a->0}
\end{align}
Here, $\langle {T}_{\mu\nu}(t,x) \rangle_{{\rm lat}}$ is 
the expectation value obtained on the lattice with $N_\tau$.
One has to determine  $b_{\mu \nu} (t)$ 
for each $t$ independently.
Then, we  take $t\to0$ extrapolation by fitting the
continuum extrapolated result 
\begin{align} 
\langle {T}_{\mu\nu}(t,x) \rangle_{\rm cont}
  = \langle {T}_{\mu\nu}(x) \rangle
      + C_{\mu \nu} t ,
  \label{eq:t->0}
\end{align}
according to Eq.~(\ref{eq:tildeTt}).
$C_{\mu \nu}$ has in principle logarithmic $t$ dependence, 
but we treat it as a constant
in our extrapolation. 

\section{Numerical results}
\label{sec:num}

\subsection{$\langle {T}_{\mu\nu}(t,x)\rangle_{\rm lat}$ and its $t$ and $a$ dependences}
\label{sec:166}

 We first focus on the result for $T=1.68T_c$ 
to see the $t$ and $a$ dependences of the numerical 
results. 
 Shown in Fig.~\ref{fig:166} are the $t$ 
dependence of $\Delta/T^4=(\epsilon -3p)/T^4$ (left) and 
entropy density $s/T^3=(\varepsilon+p)/T^4$ (right) as functions of $tT^2$
at  fixed temperature, $T/T_c=1.68$, for three different values of the lattice spacing,
$\beta=6.719$, $6.941$ and $7.117$ 
 ($a=0.033$, $0.025$ and $0.020\,\mathrm{fm}$).
 For $\Delta/T^4$, the improved operator in Eq.~(\ref{eq:E_imp}) is adopted.
 Let us discuss the three region of $t$ separately:
 (i) For $0< \sqrt{8t}\lesssim a$,  the lattice discretization effect 
 becomes prominent as discussed in Sec.~\ref{sec:ep}.
 One finds, particularly in the right panel,  that this region becomes
 narrower as $a$ decreases.
(ii) For the smallest $a$ in this figure (red points),
 $\Delta/T^4$ has a plateau and $s/T^3$ has a linear behavior
   in the range $0.005\lesssim tT^2 \lesssim 0.015$ in accordance with 
Eq.~(\ref{eq:tildeTt}). 
(iii) The deviation from the linear behavior is seen for 
  $tT^2\gtrsim 0.015$, which is attributed to the over-smearing 
as discussed in Sec.~\ref{sec:ep}. 
These consideration indicates that there exists a window of $t$
from which 
the values of $\Delta/T^4$ and~$s/T^3$ at $t=0$
can be extracted.

\begin{figure*}
  \centering
  \includegraphics[width=0.45\textwidth,clip]{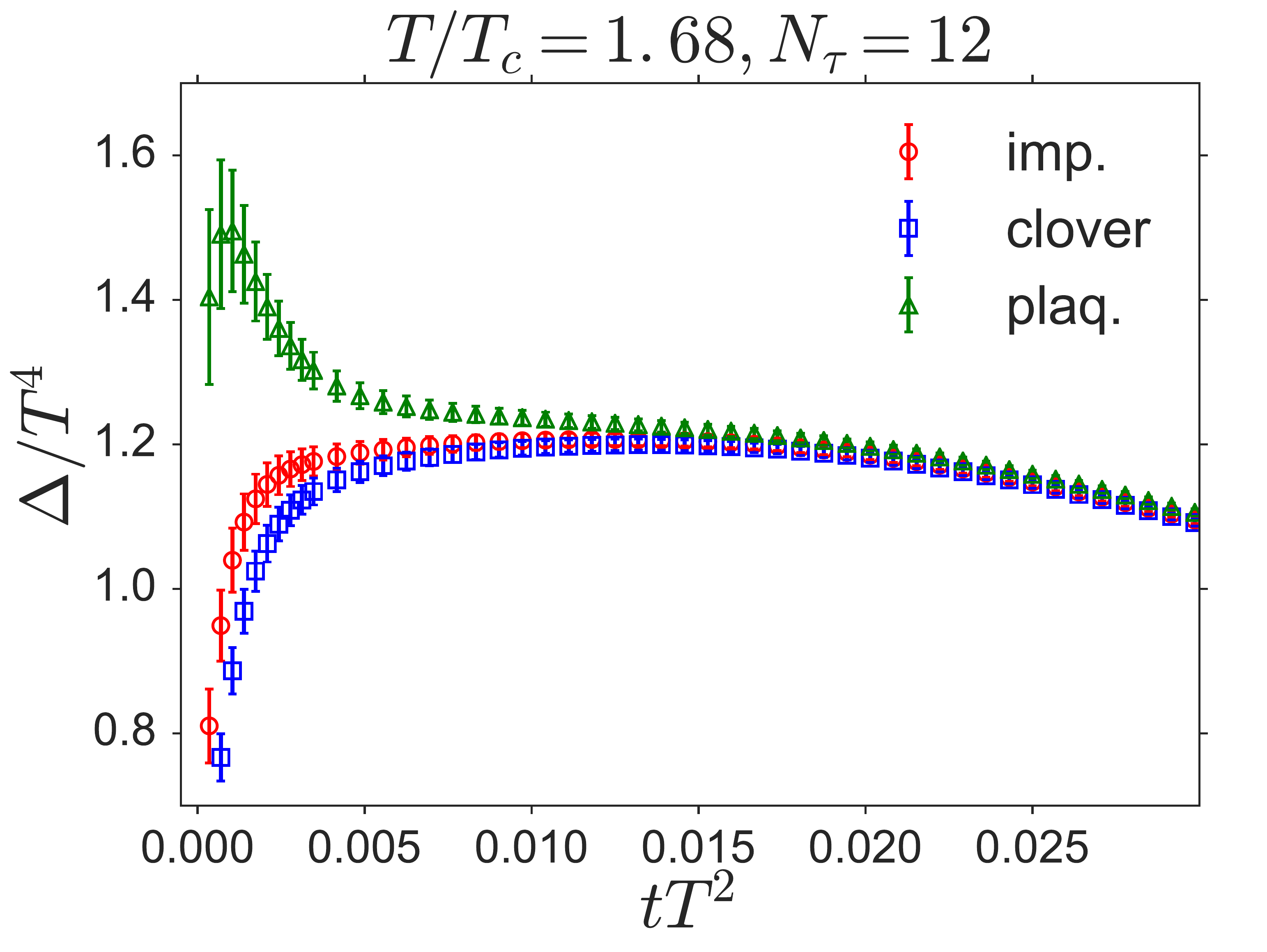}
  \includegraphics[width=0.45\textwidth,clip]{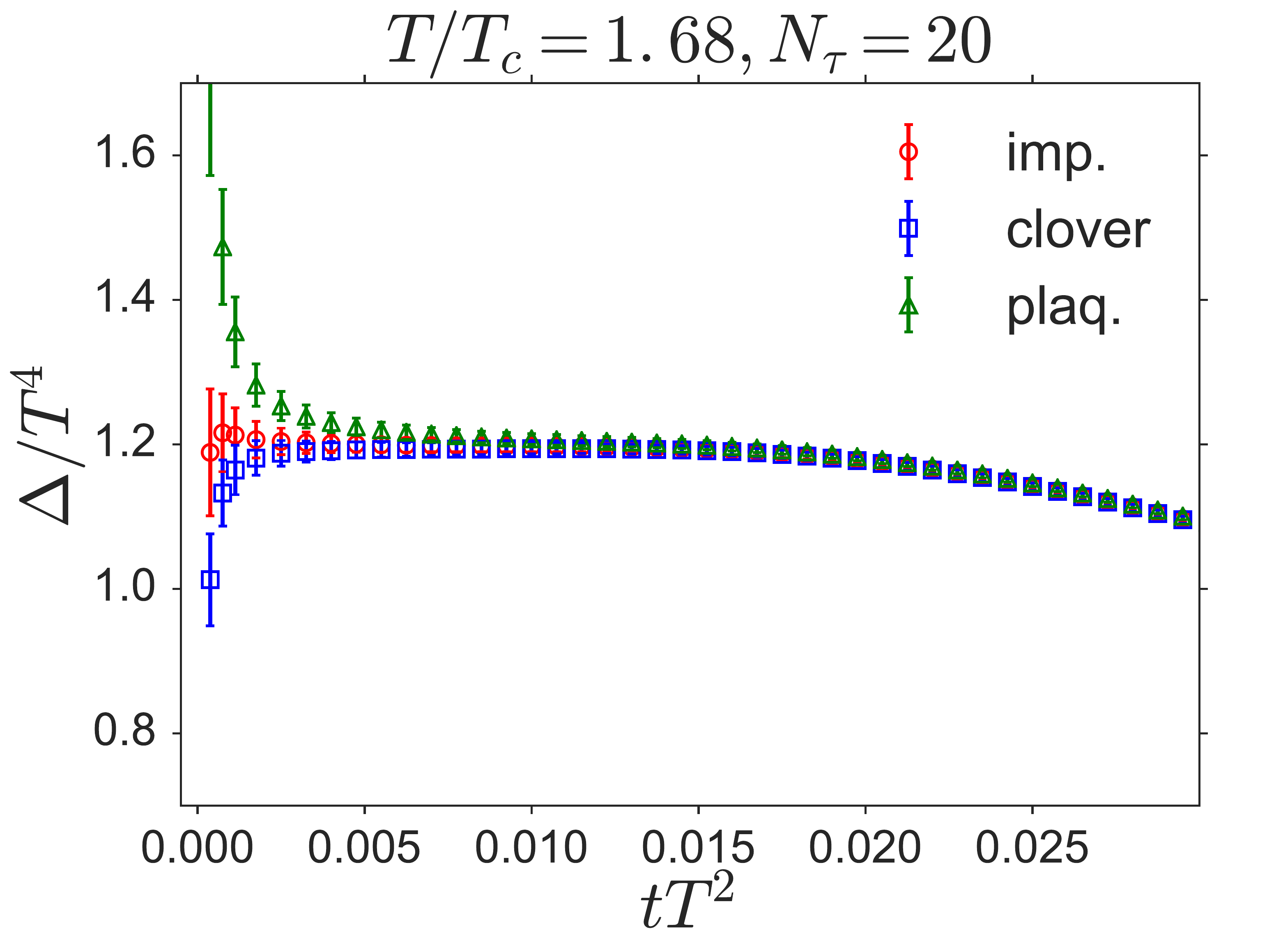}
\caption{
  $t$ dependence of $\Delta/T^4$ for 
  $T/T_c=1.68$ 
  with $N_\tau=12$ (left) and $N_\tau=20$ (right)
  calculated by different discretizations, 
  $E(t,x)_{\rm imp}$, $E(t,x)_{\rm clover}$ and $E(t,x)_{\rm plaq}$.
}
\label{fig:improve}
\end{figure*}

To check the effect of different choices for the   
operator $E(t,x)$ in  $\Delta/T^4$,
we compare three cases in Fig.~\ref{fig:improve},  $E(t,x)_{\rm imp}$, $E(t,x)_{\rm clover}$ and $E(t,x)_{\rm plaq}$,
for $T/T_c=1.68$ and $N_{\tau}=12$, $20$.  
In both figures, the improved operator Eq.~(\ref{eq:E_imp}) 
 shows least discretization error for~$\Delta/T^4$.

\subsection{Double extrapolation}

\begin{figure*}
  \centering
  \includegraphics[width=0.4\textwidth,clip]{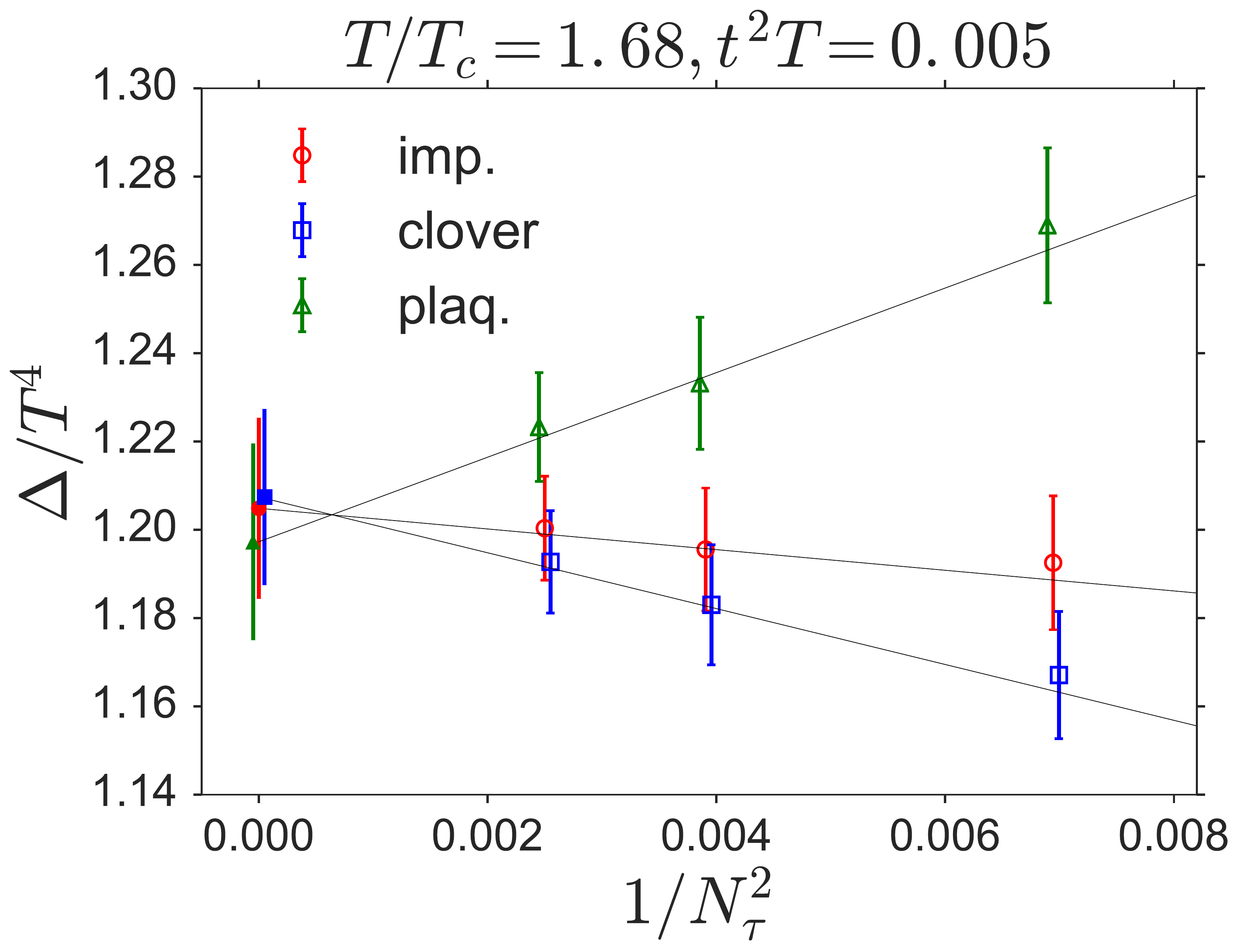}
  \includegraphics[width=0.4\textwidth,clip]{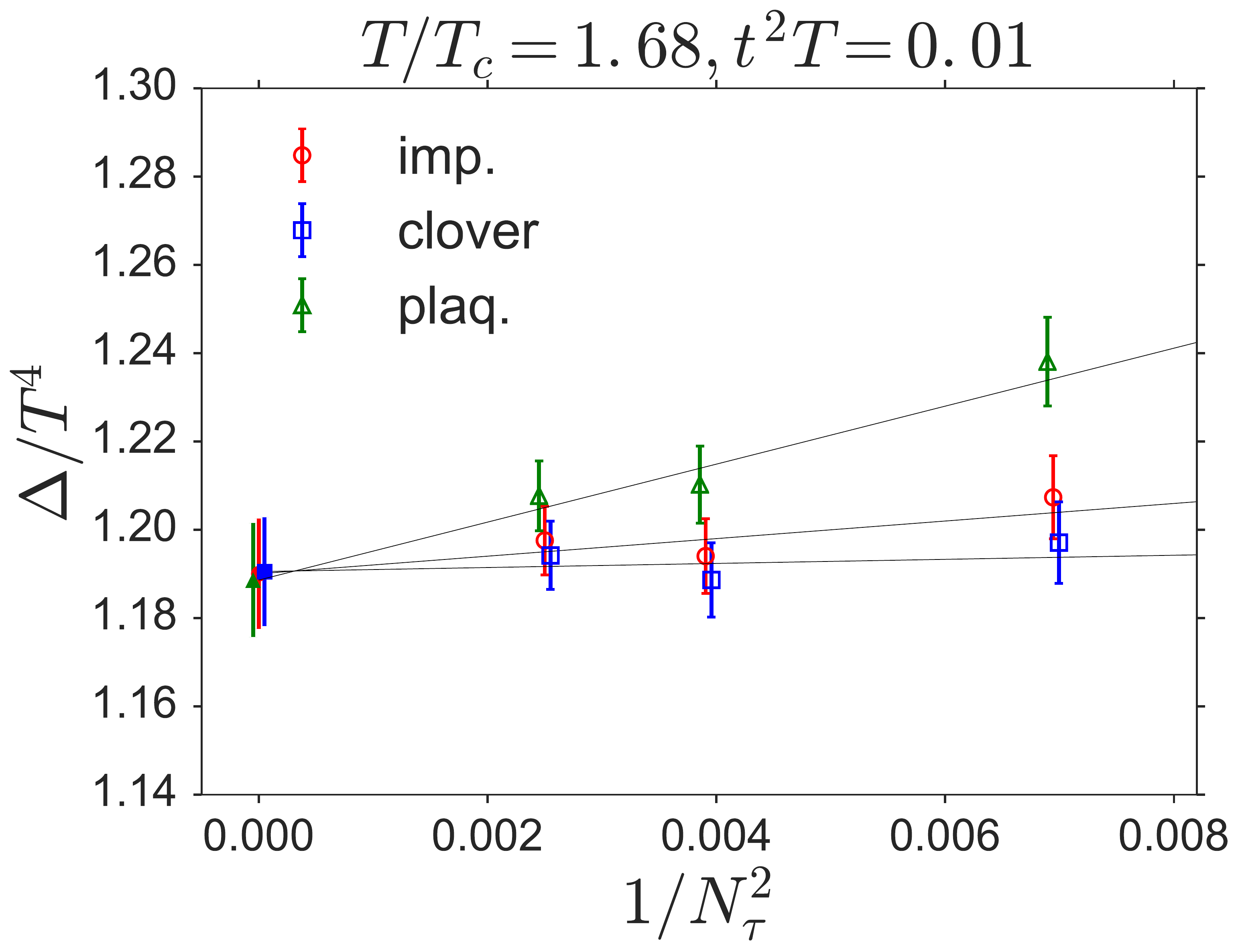}

  \includegraphics[width=0.4\textwidth,clip]{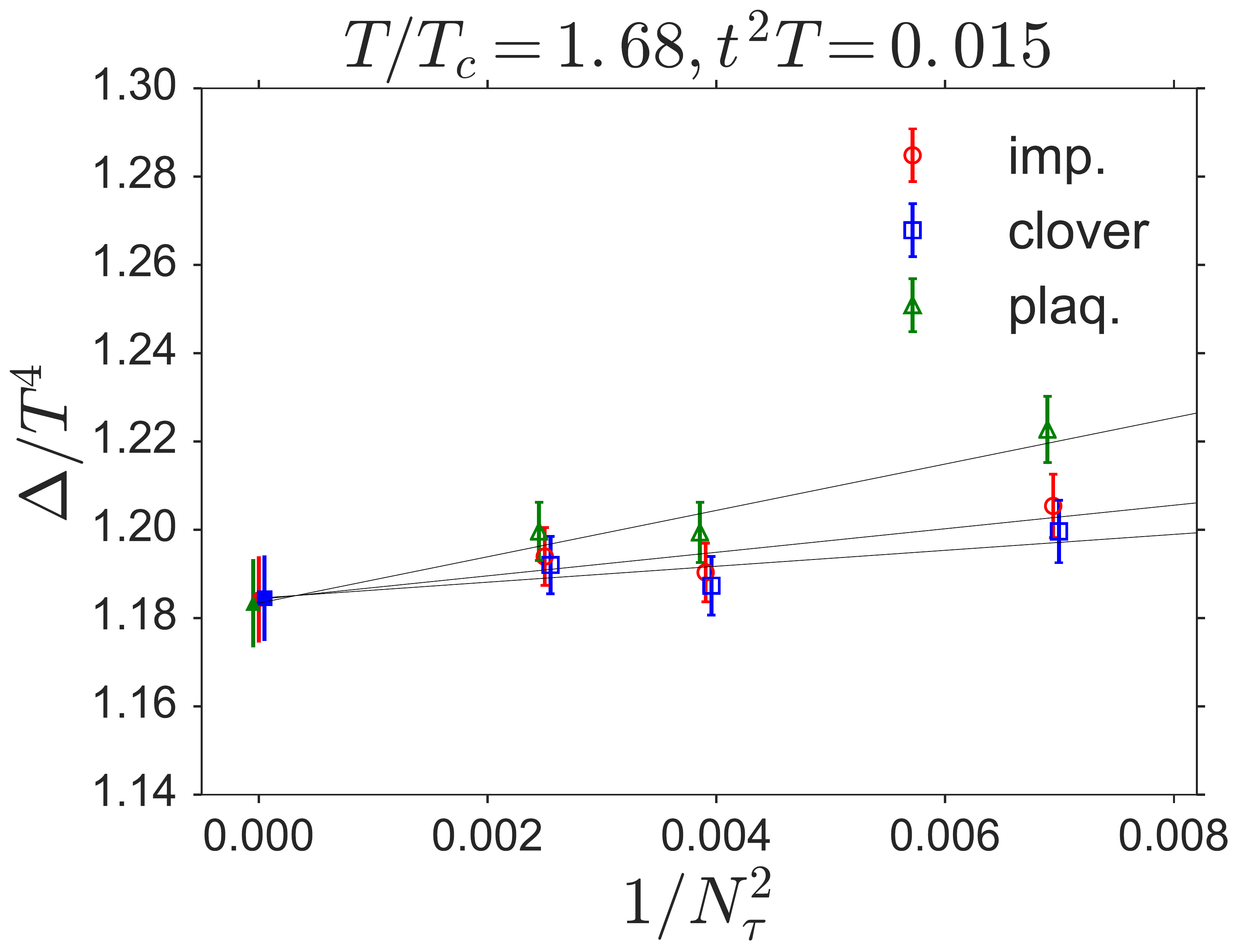}
  \includegraphics[width=0.4\textwidth,clip]{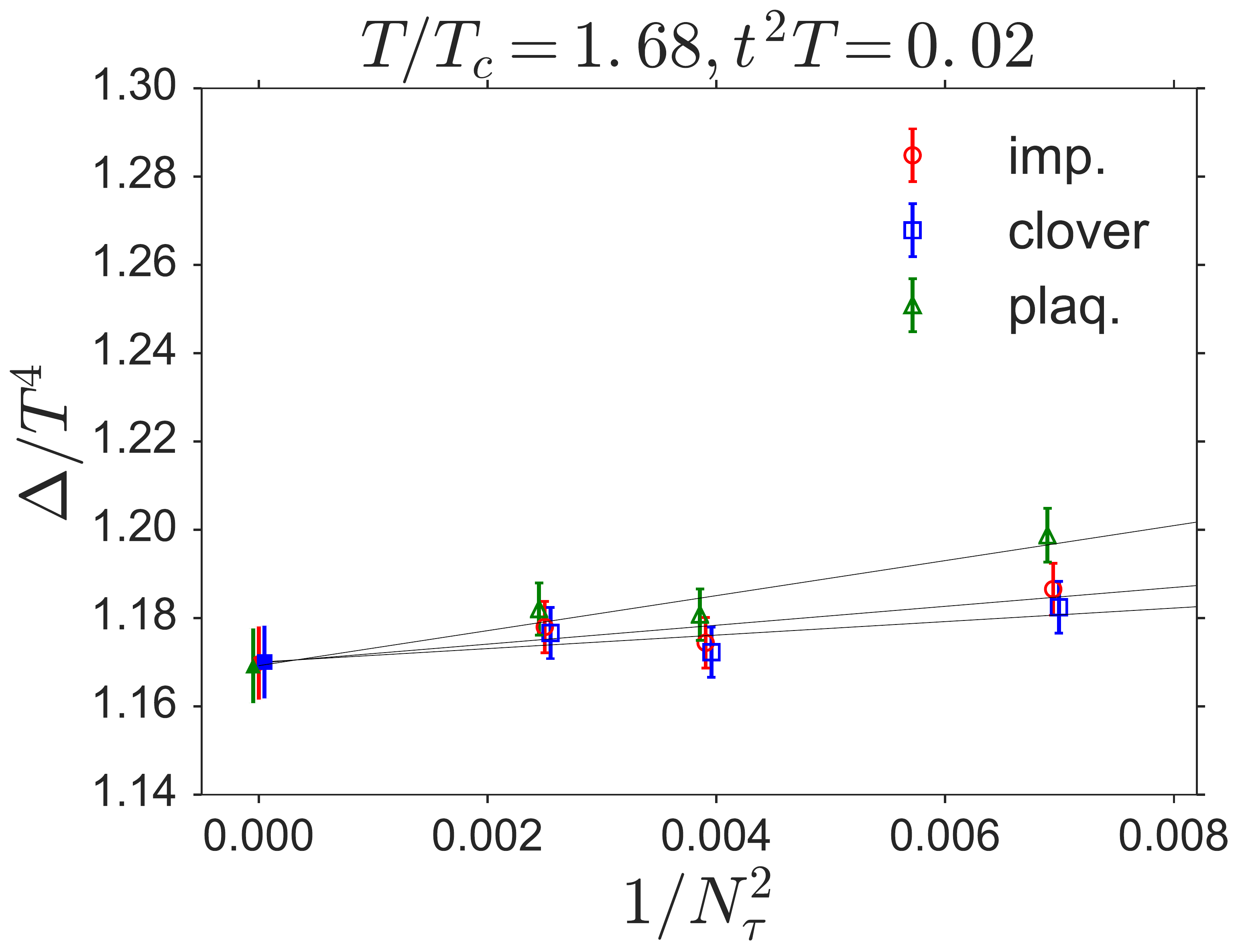}
  \caption{
    $N_\tau$ dependence of $\Delta/T^4$ at $tT^2 = 0.005$, $0.01$, $0.015$ and $0.02$
    together with the result of continuum extrapolation
    using Eq.~(\ref{eq:a->0}).
    The results with three discretizations for $E(t,x)$ are plotted.
    \label{fig:cont_limit_anomaly}
  }
\end{figure*}

\begin{figure*}
  \centering
  \includegraphics[width=0.4\textwidth,clip]{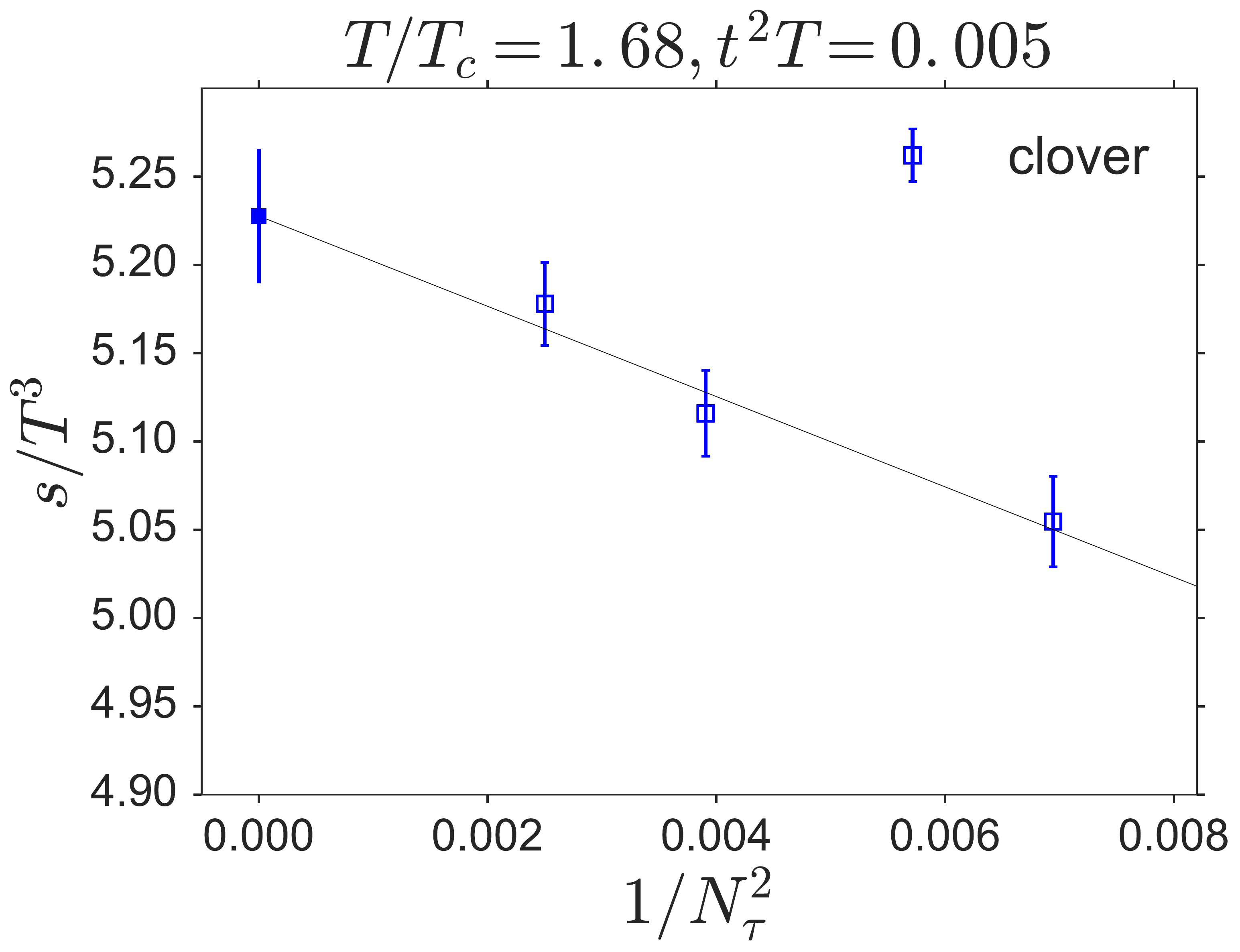}
  \includegraphics[width=0.4\textwidth,clip]{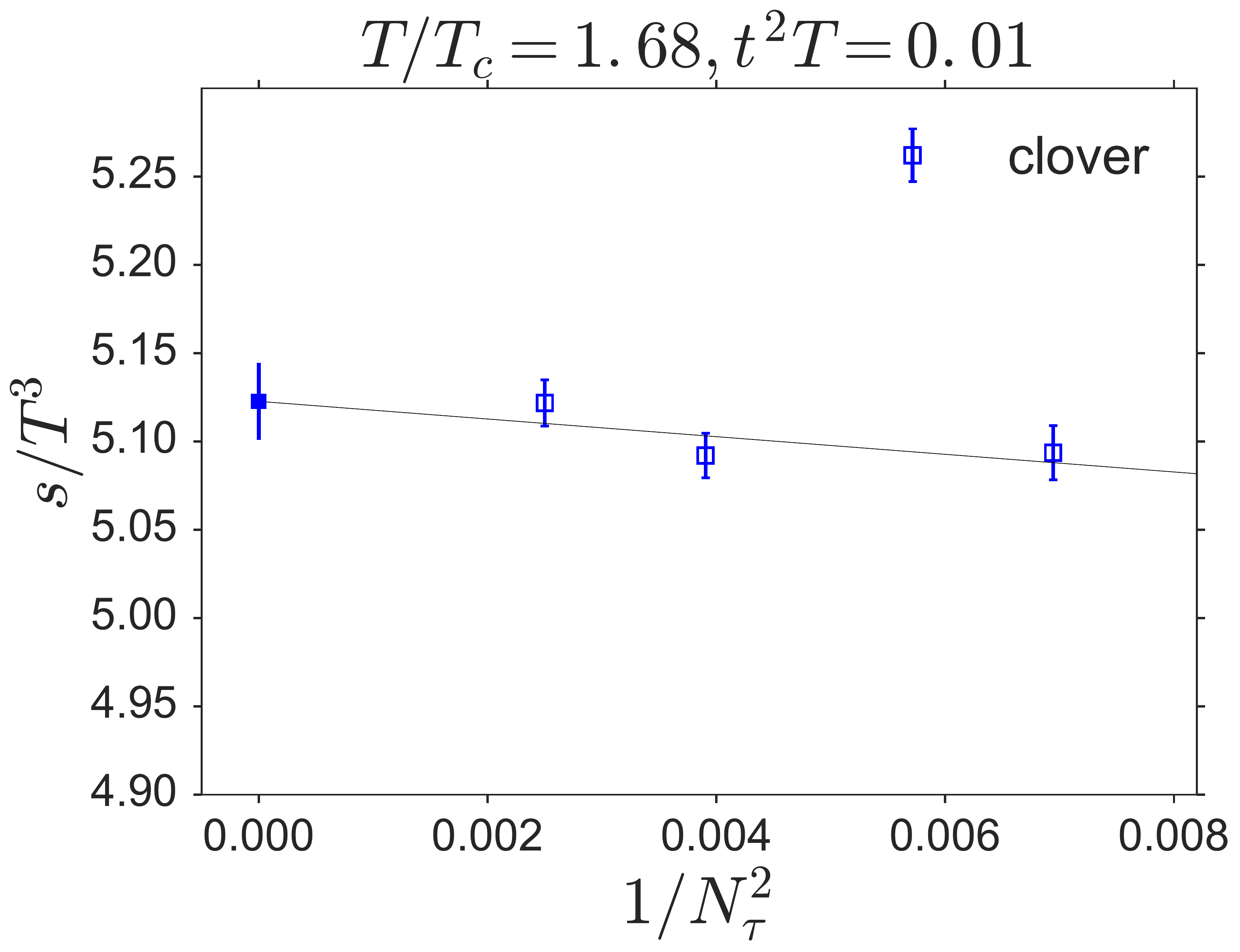}

  \includegraphics[width=0.4\textwidth,clip]{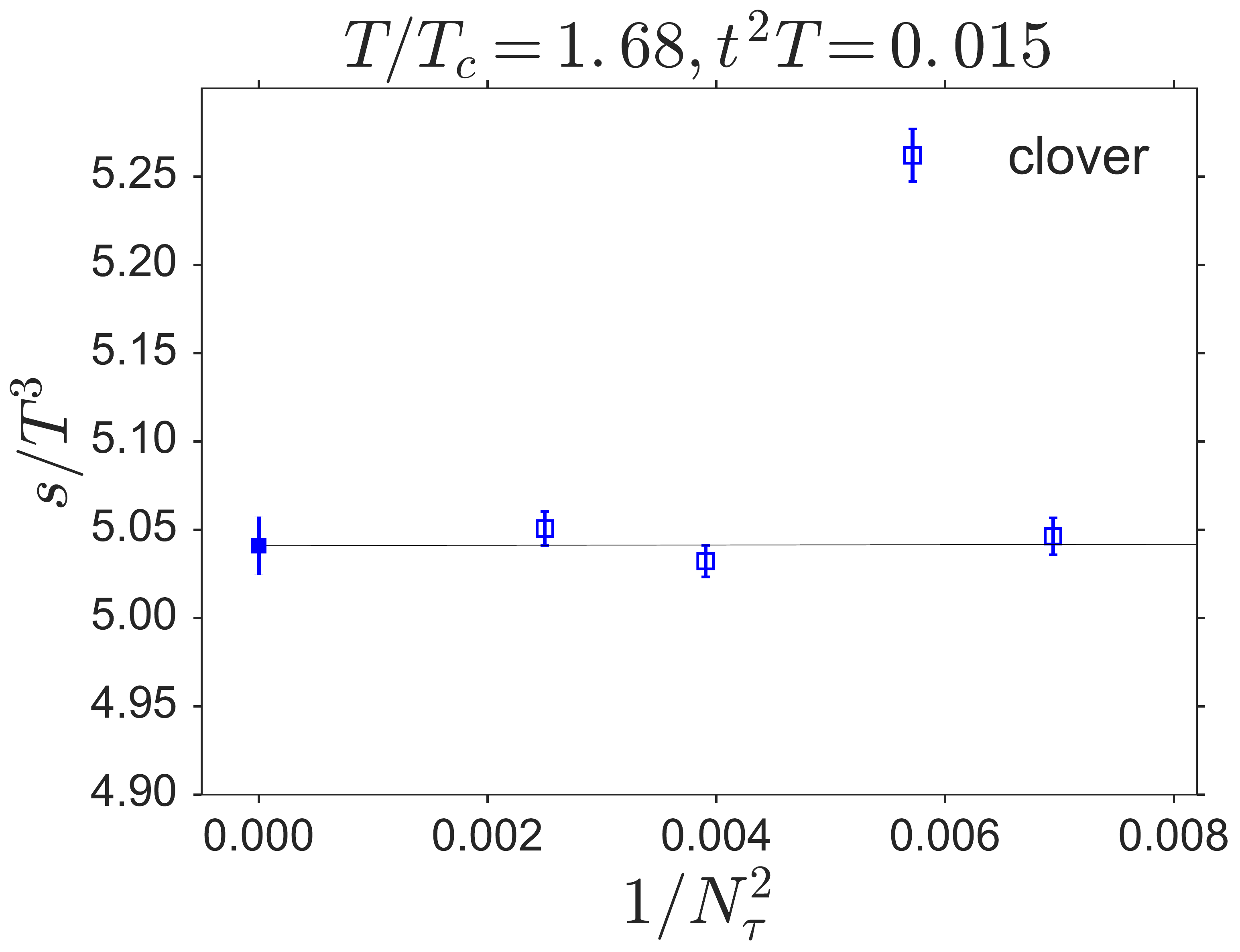}
  \includegraphics[width=0.4\textwidth,clip]{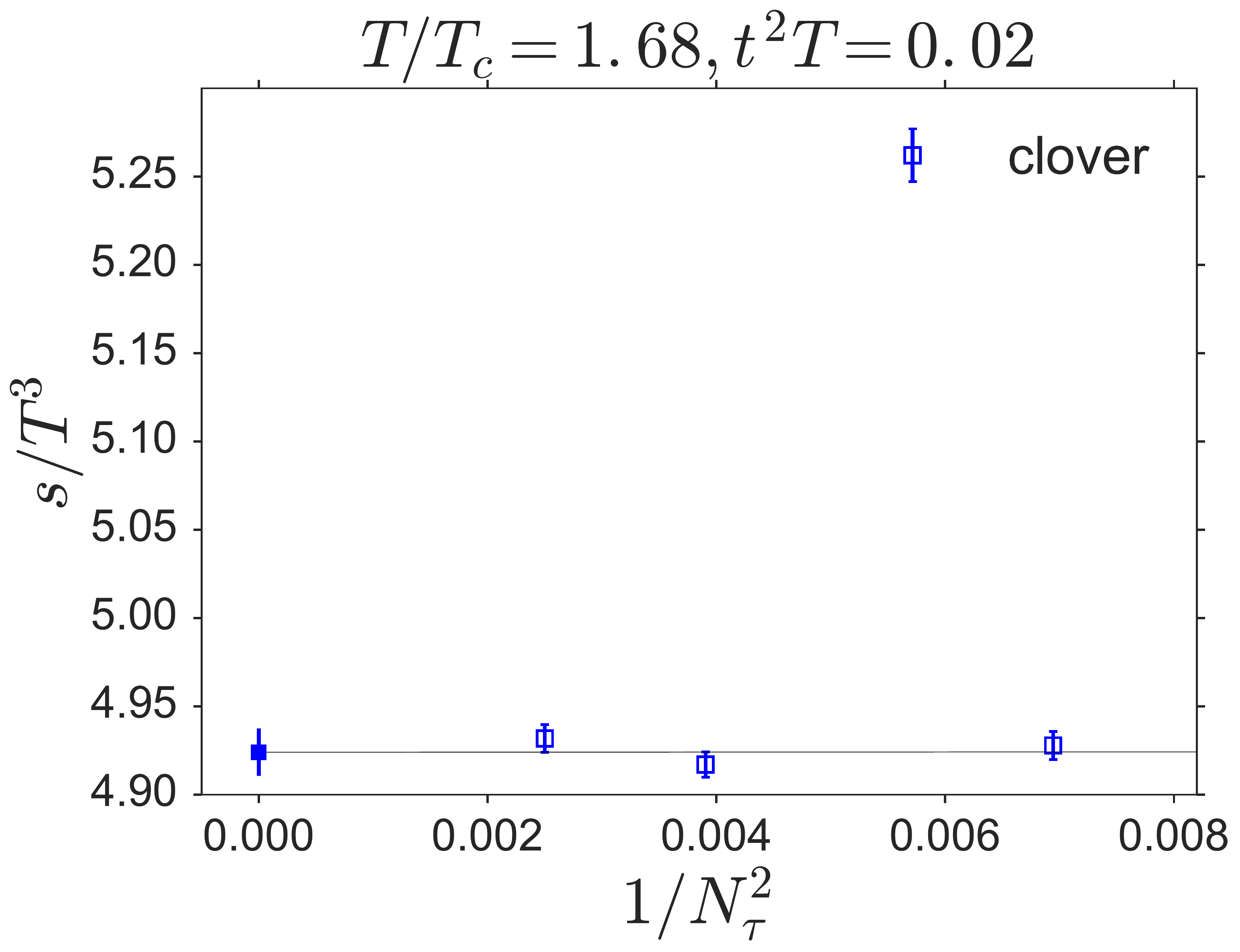}
  \caption{
    $N_\tau$ dependence of $s/T^3$ at $tT^2 = 0.005$, $0.01$, $0.015$ and $0.02$
    together with the result of continuum extrapolation
    using Eq.~(\ref{eq:a->0}).
    \label{fig:cont_limit_entropy}
  }
\end{figure*}

\begin{figure*}
  \includegraphics[width=0.45\textwidth,clip]{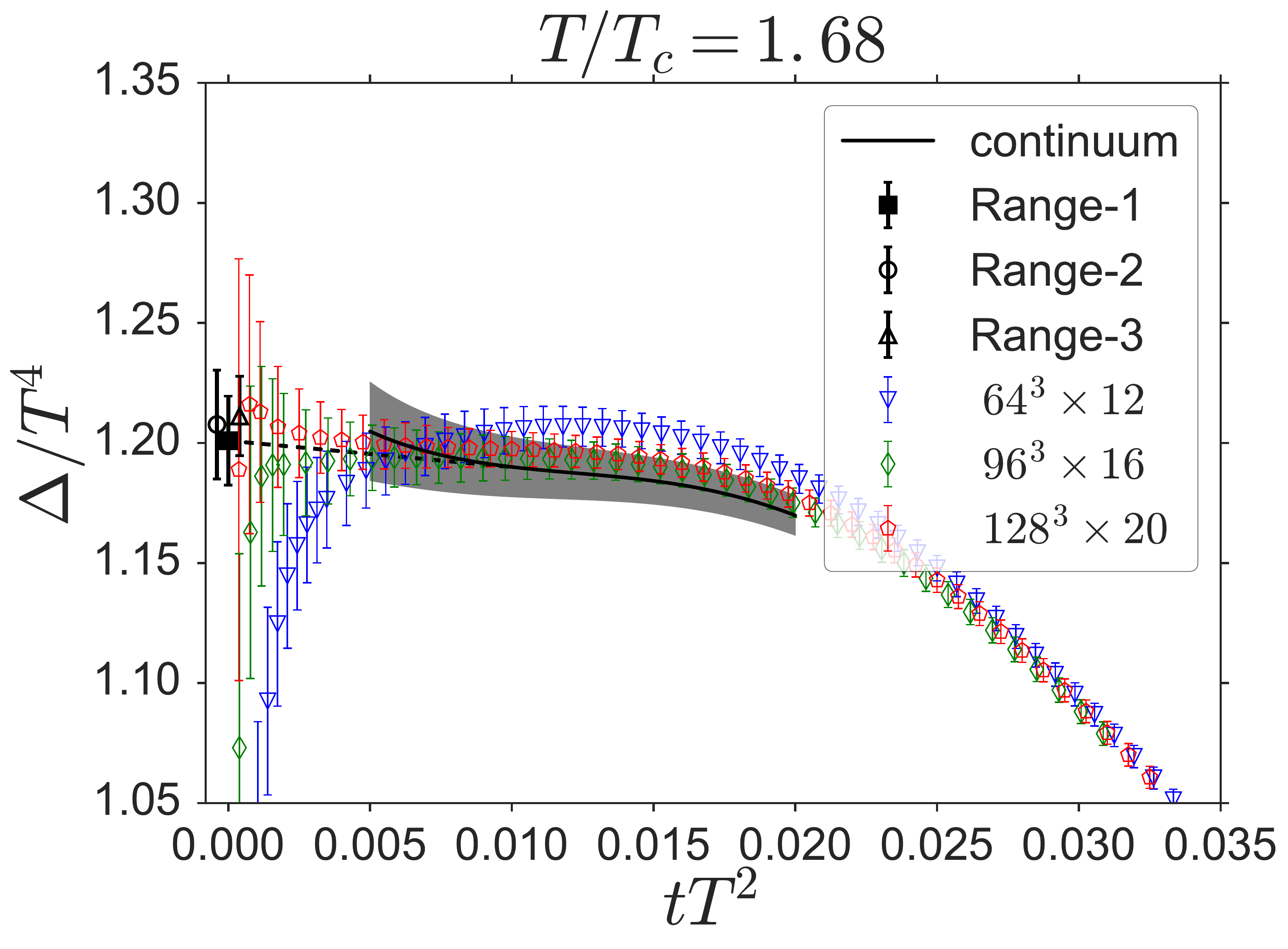}
  \includegraphics[width=0.45\textwidth,clip]{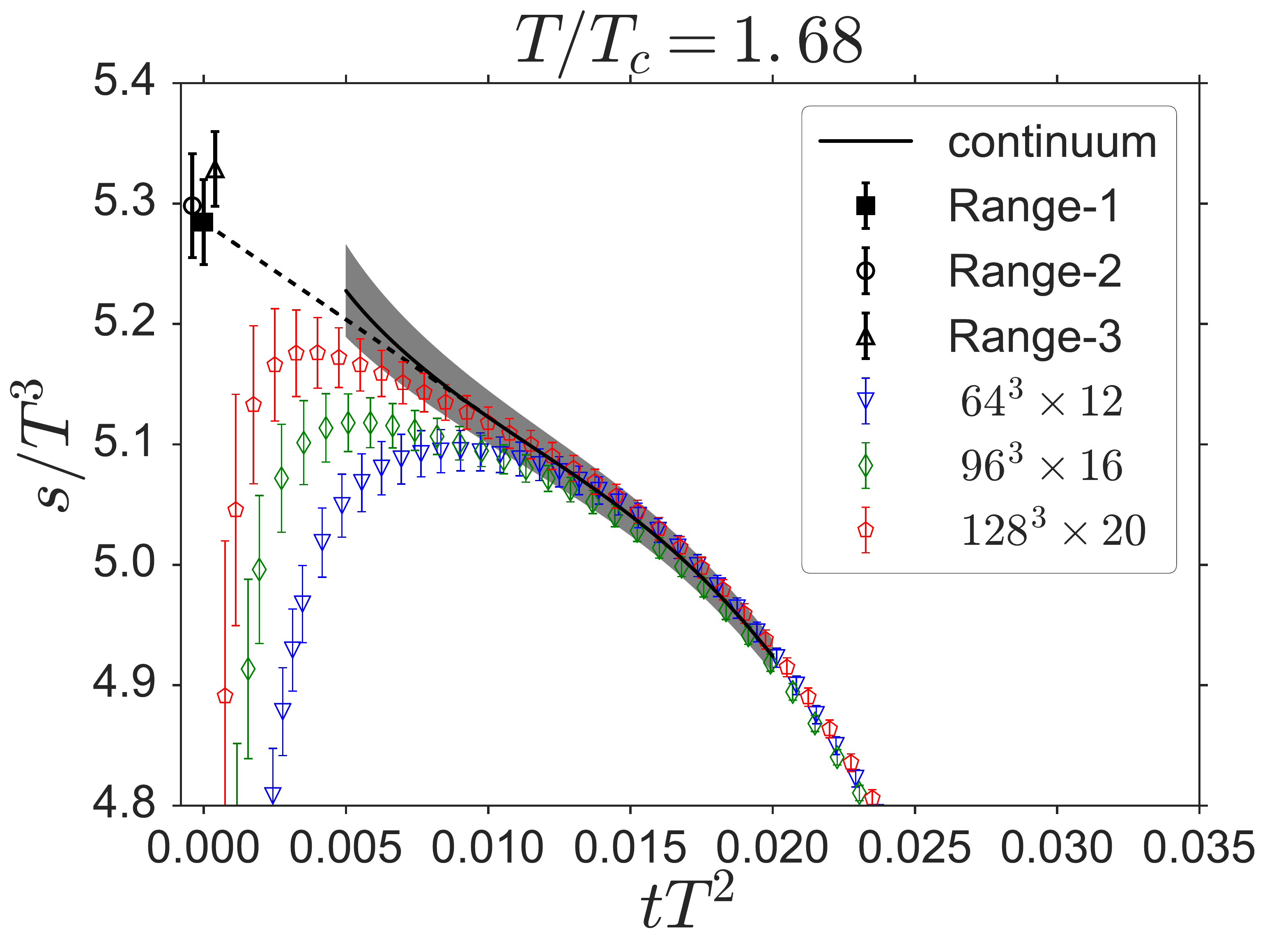}
\caption{
Results of continuum extrapolation (black band)
  for $\Delta/T^4$ (left) and $s/T^3$ (right) as functions of $tT^2$. 
  The extrapolation to $t=0$ using the data in Range-1
  is shown by the dashed line, and the extrapolated value with the error
  is given by the filled square at $t=0$.
  The extrapolated values with Range-2 and Range-3 are also shown
  around the origin.
  }  
\label{fig:166cont}
\end{figure*}

\begin{figure}
  \centering
  \includegraphics[width=0.235\textwidth,clip]{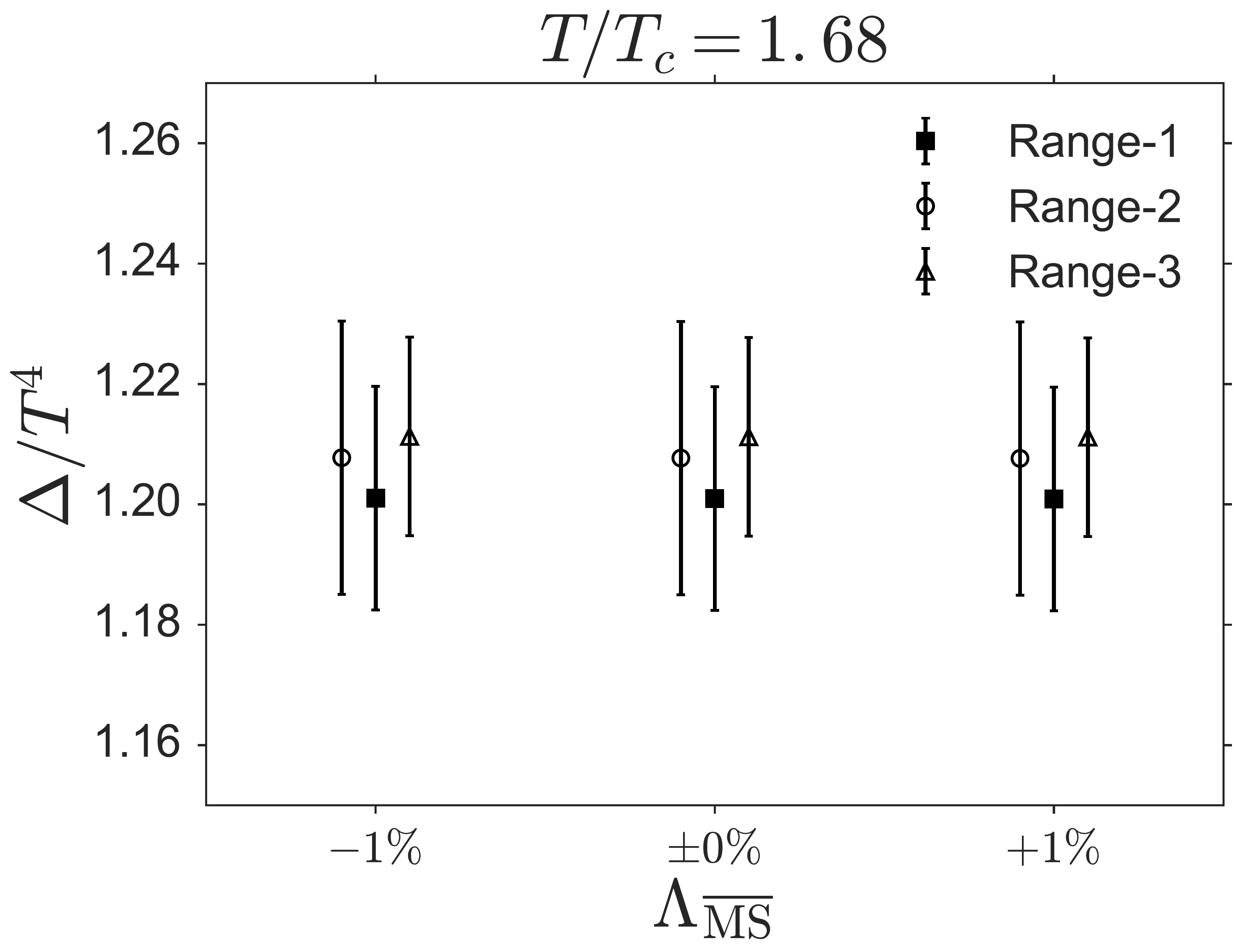}
  \includegraphics[width=0.235\textwidth,clip]{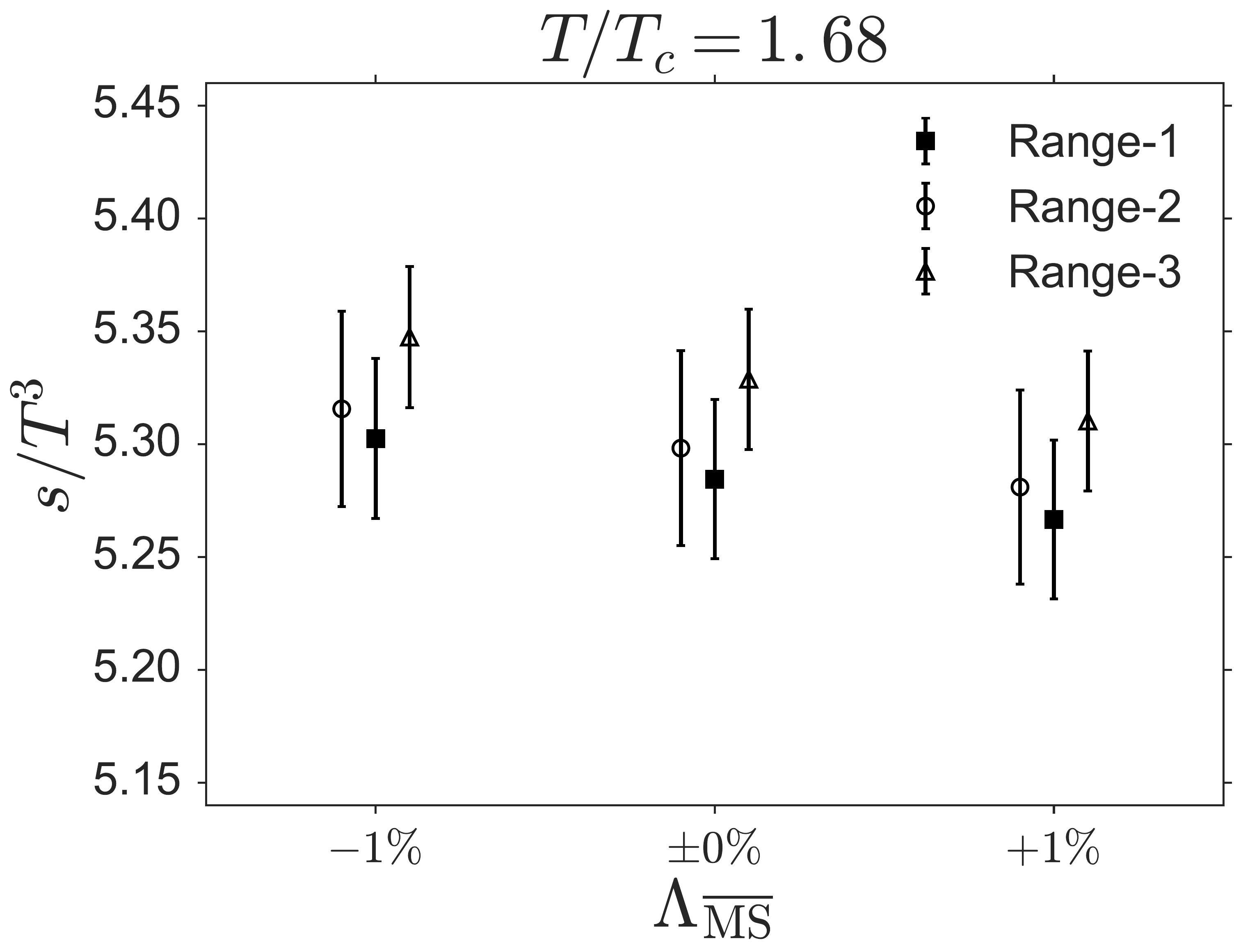}
  \caption{
    Systematic errors originated from $\pm 1$\% of change of $\Lambda_{\overline{\mathrm{MS}}}$.
    \label{fig:lambda_sys}
  }
\end{figure}

\begin{figure}
  \centering
  \includegraphics[width=0.235\textwidth,clip]{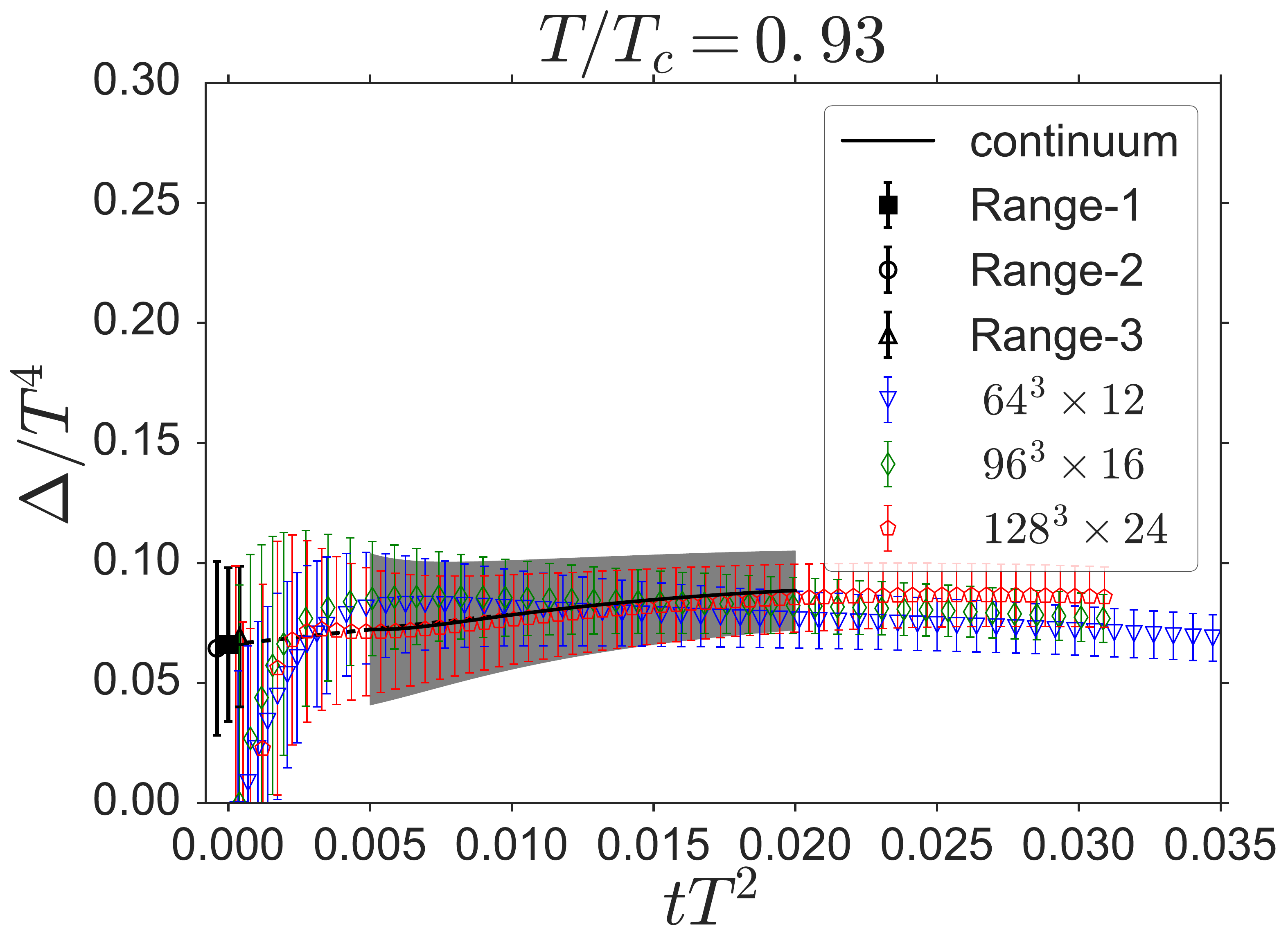}
  \includegraphics[width=0.235\textwidth,clip]{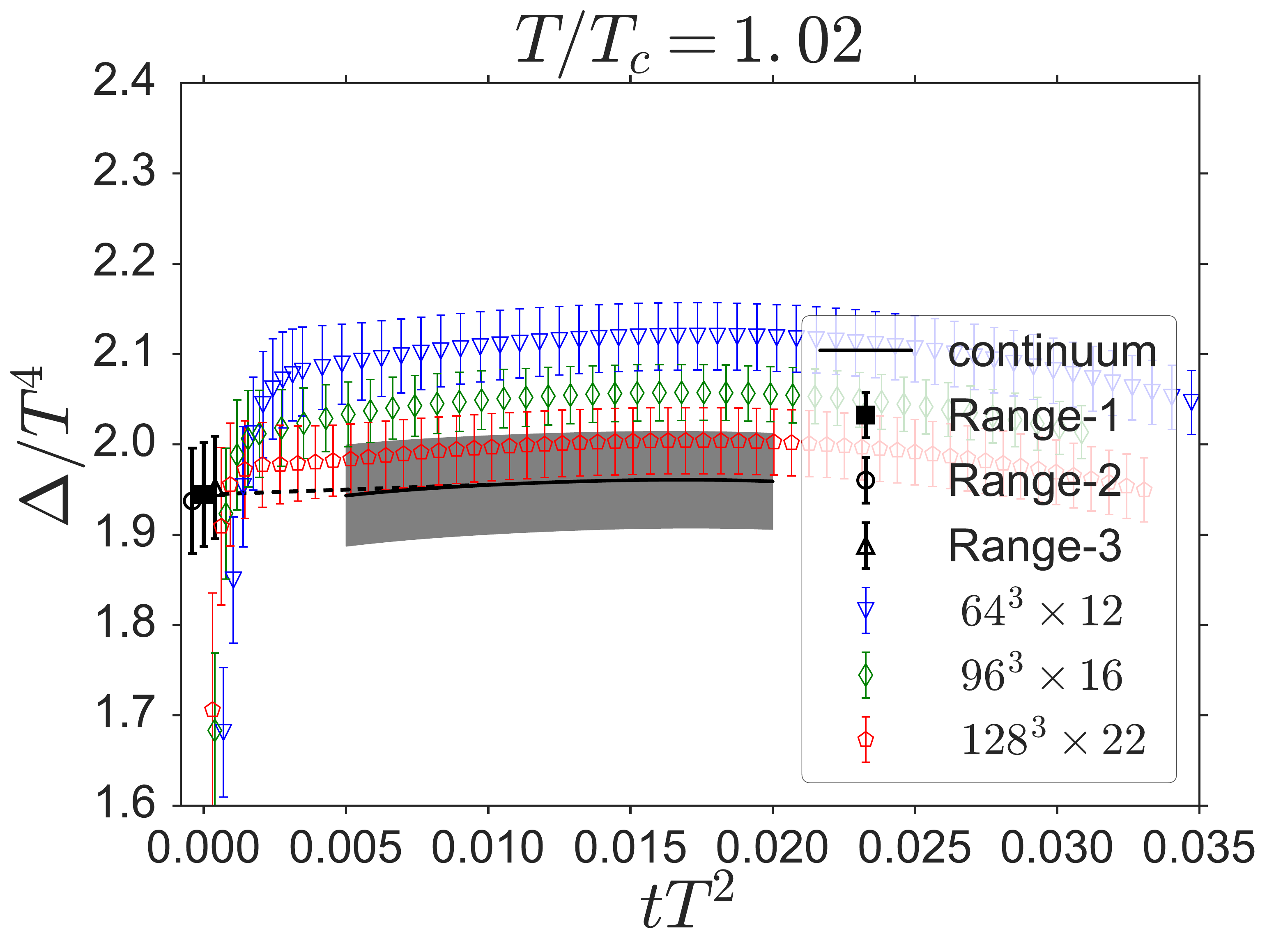}
  \includegraphics[width=0.235\textwidth,clip]{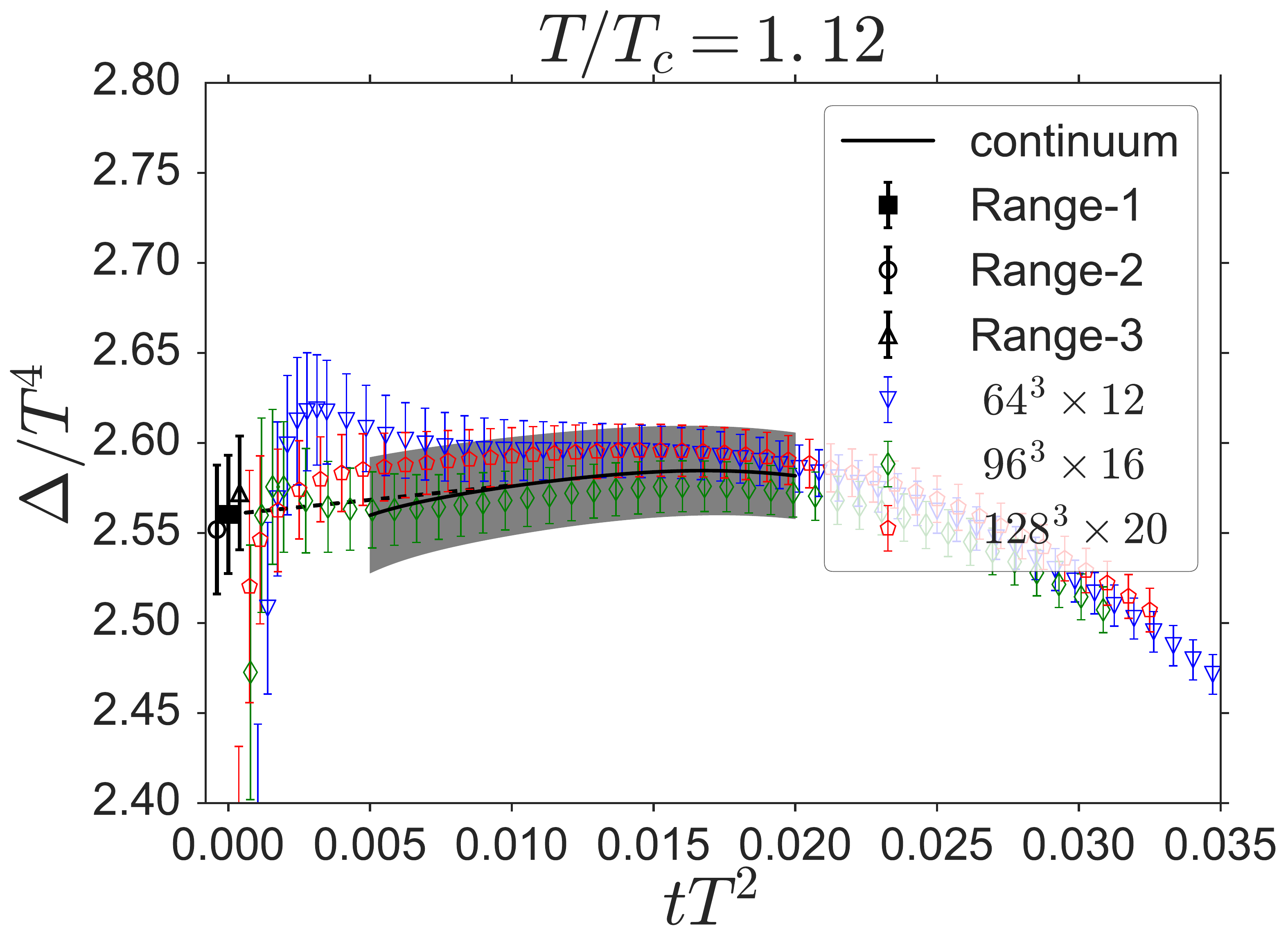}
  \includegraphics[width=0.235\textwidth,clip]{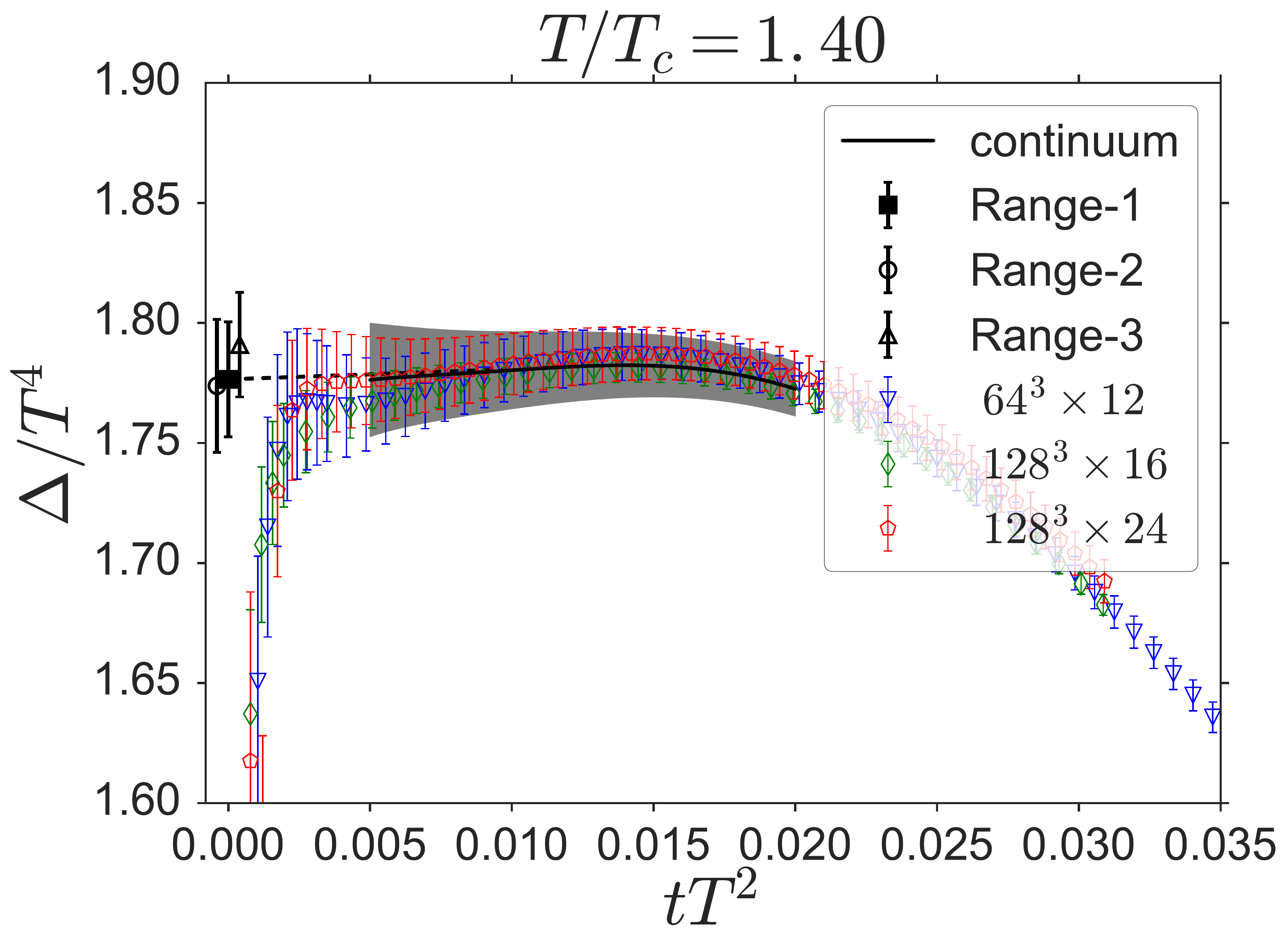}
\caption{
 Similar plots with the left panel of Fig.~\ref{fig:166cont} for different values of $T/T_c$.
}
\label{fig:e-3p}
\end{figure}

\begin{figure}
  \centering
  \includegraphics[width=0.235\textwidth,clip]{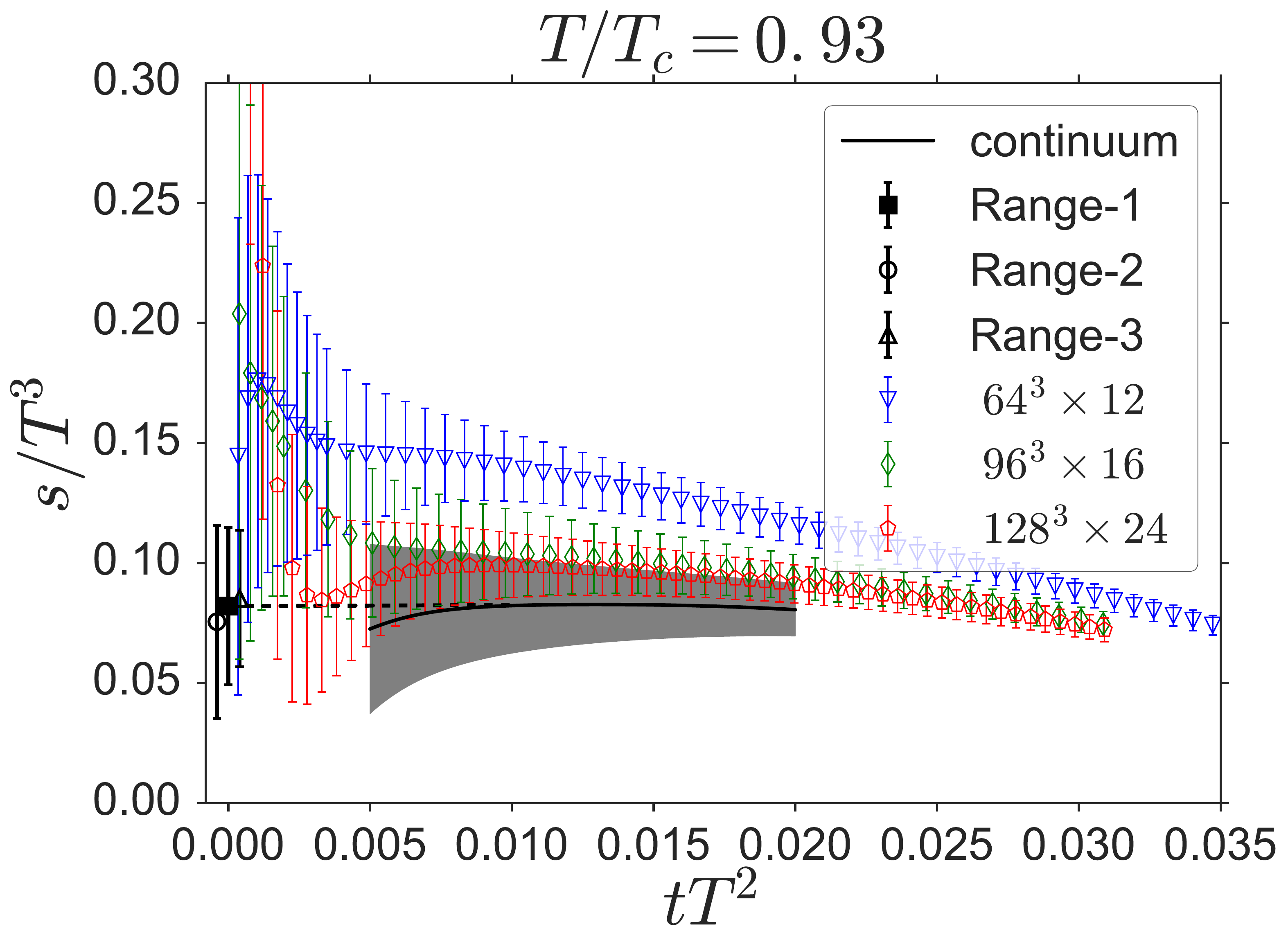}
  \includegraphics[width=0.235\textwidth,clip]{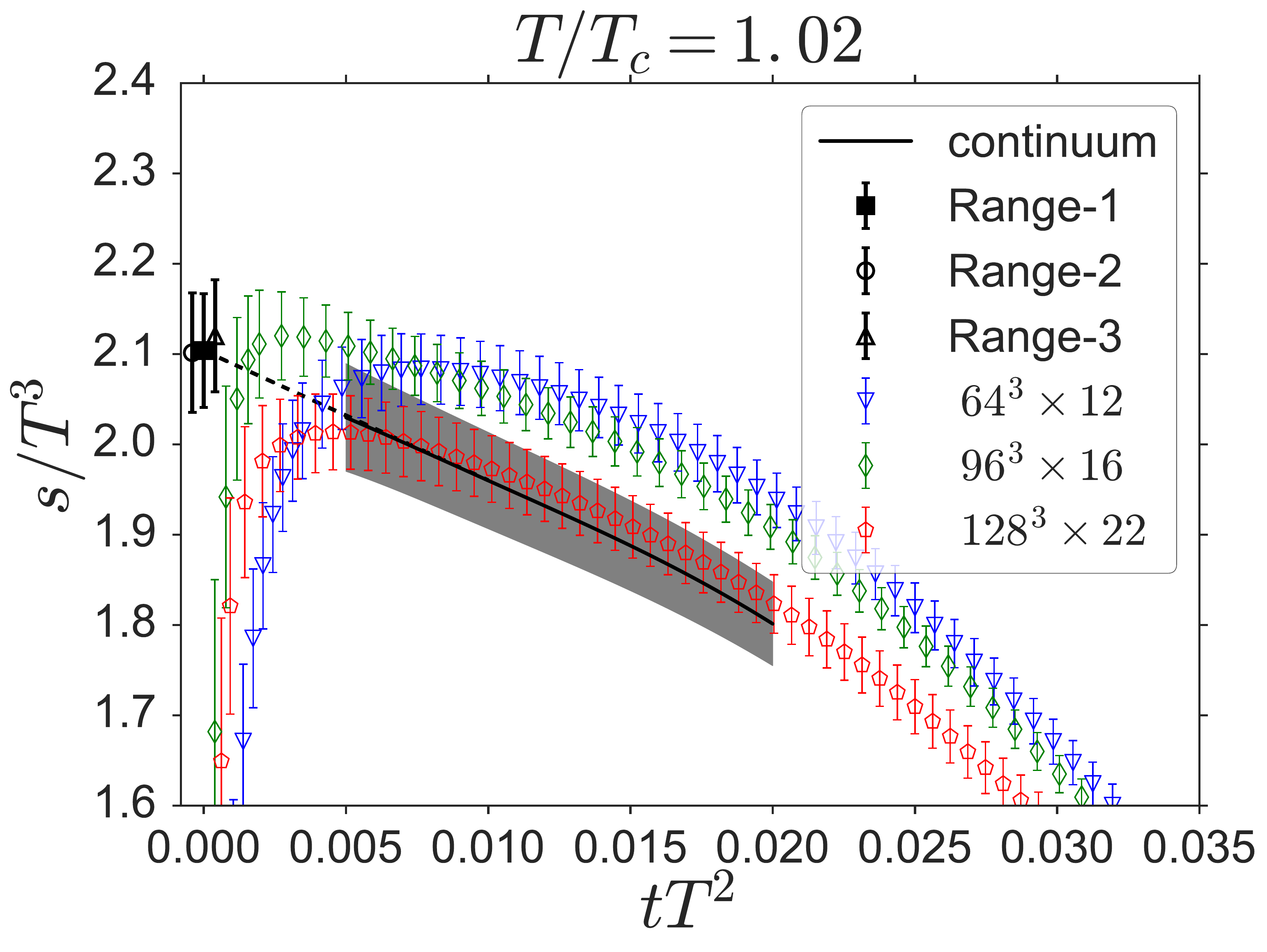}
  \includegraphics[width=0.235\textwidth,clip]{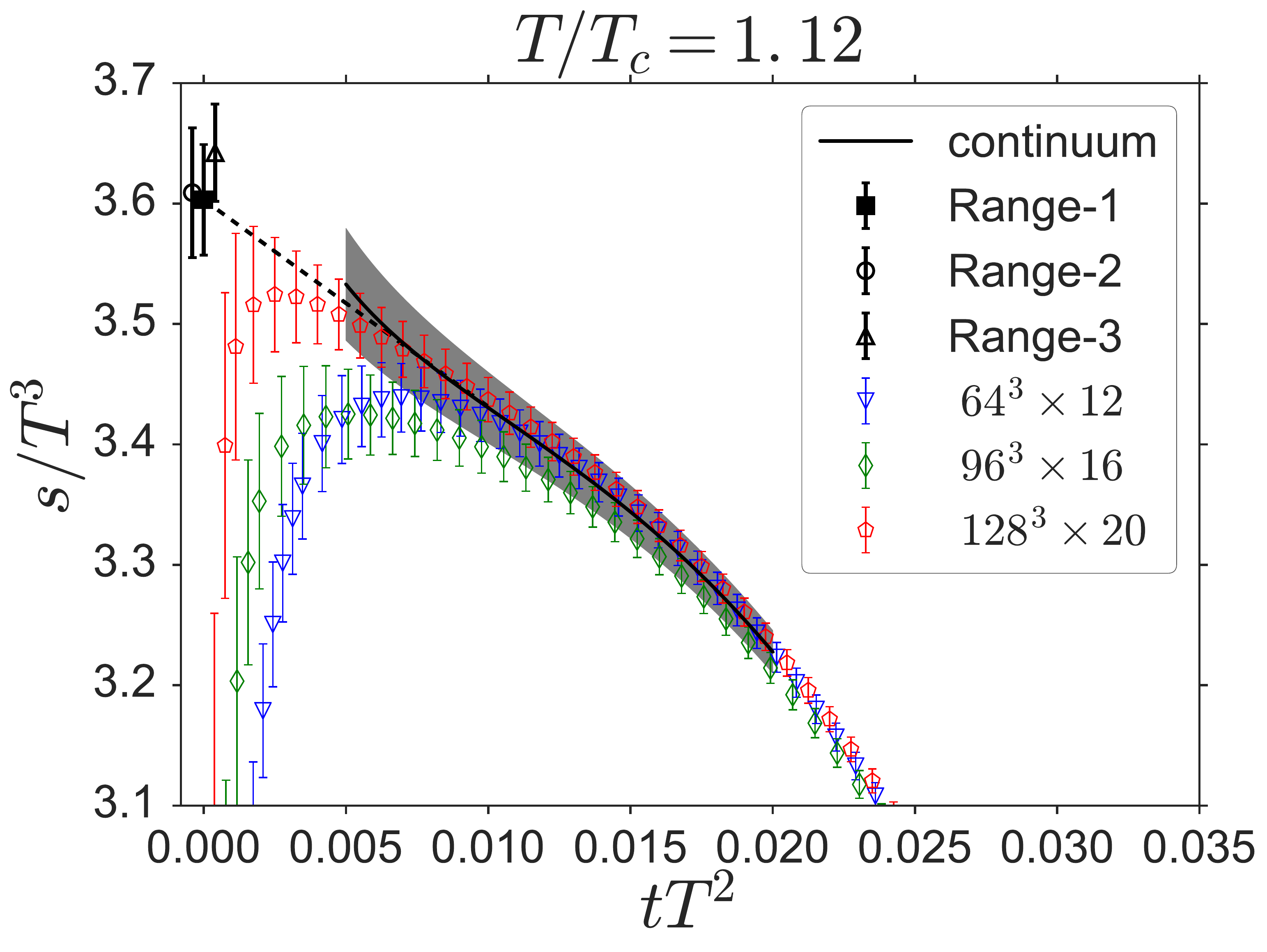}
  \includegraphics[width=0.235\textwidth,clip]{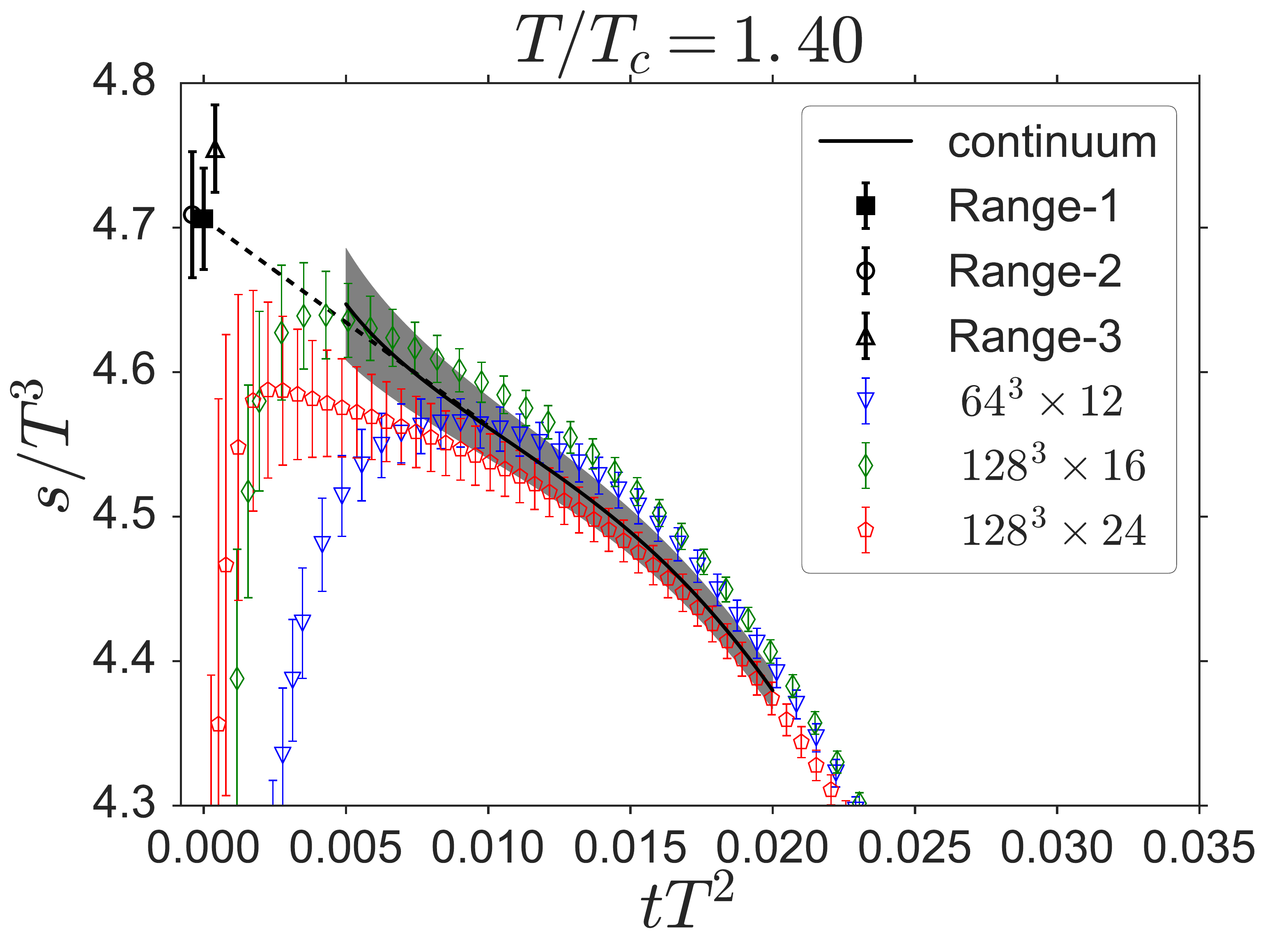}
  \includegraphics[width=0.235\textwidth,clip]{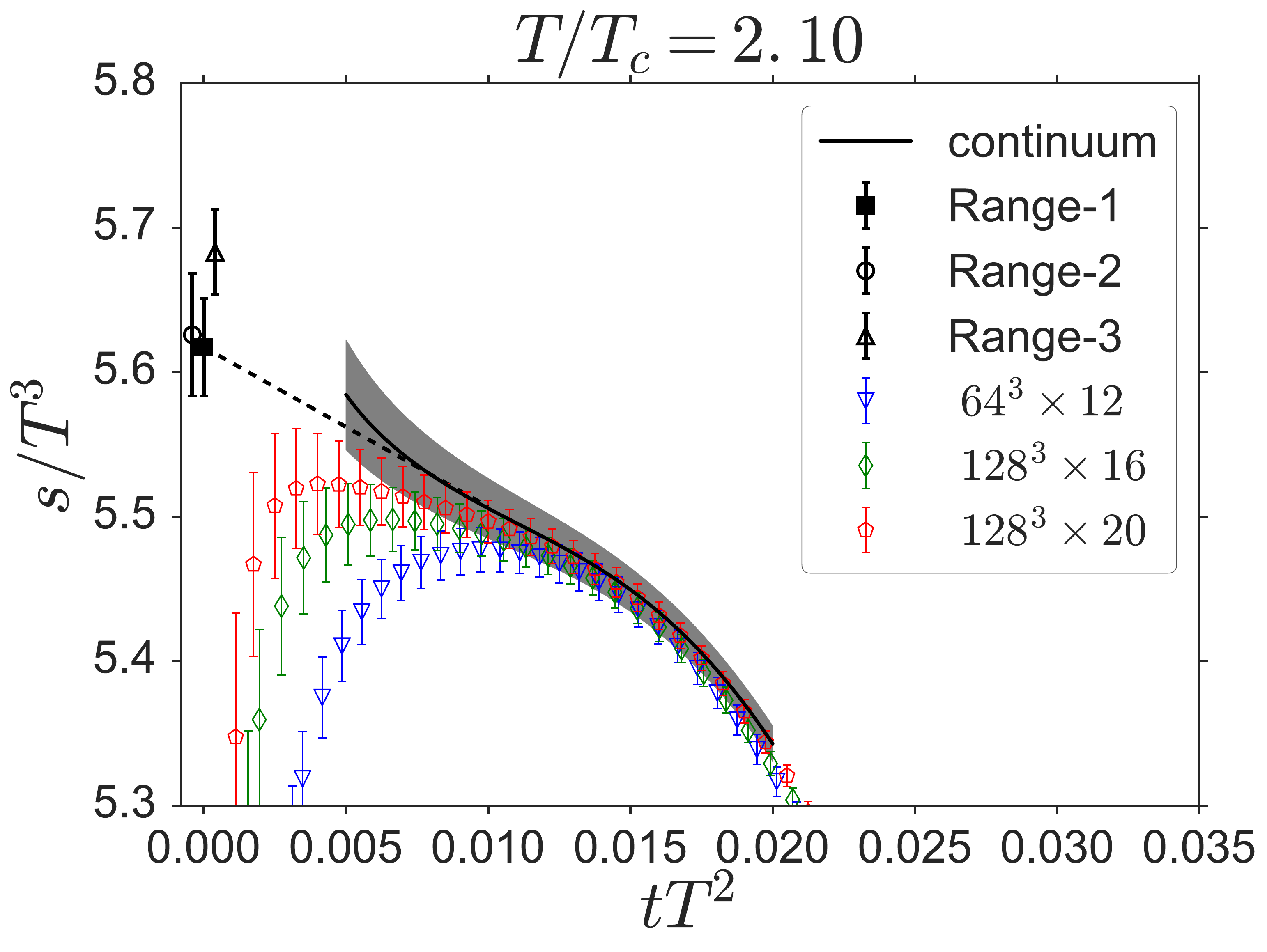}
  \includegraphics[width=0.235\textwidth,clip]{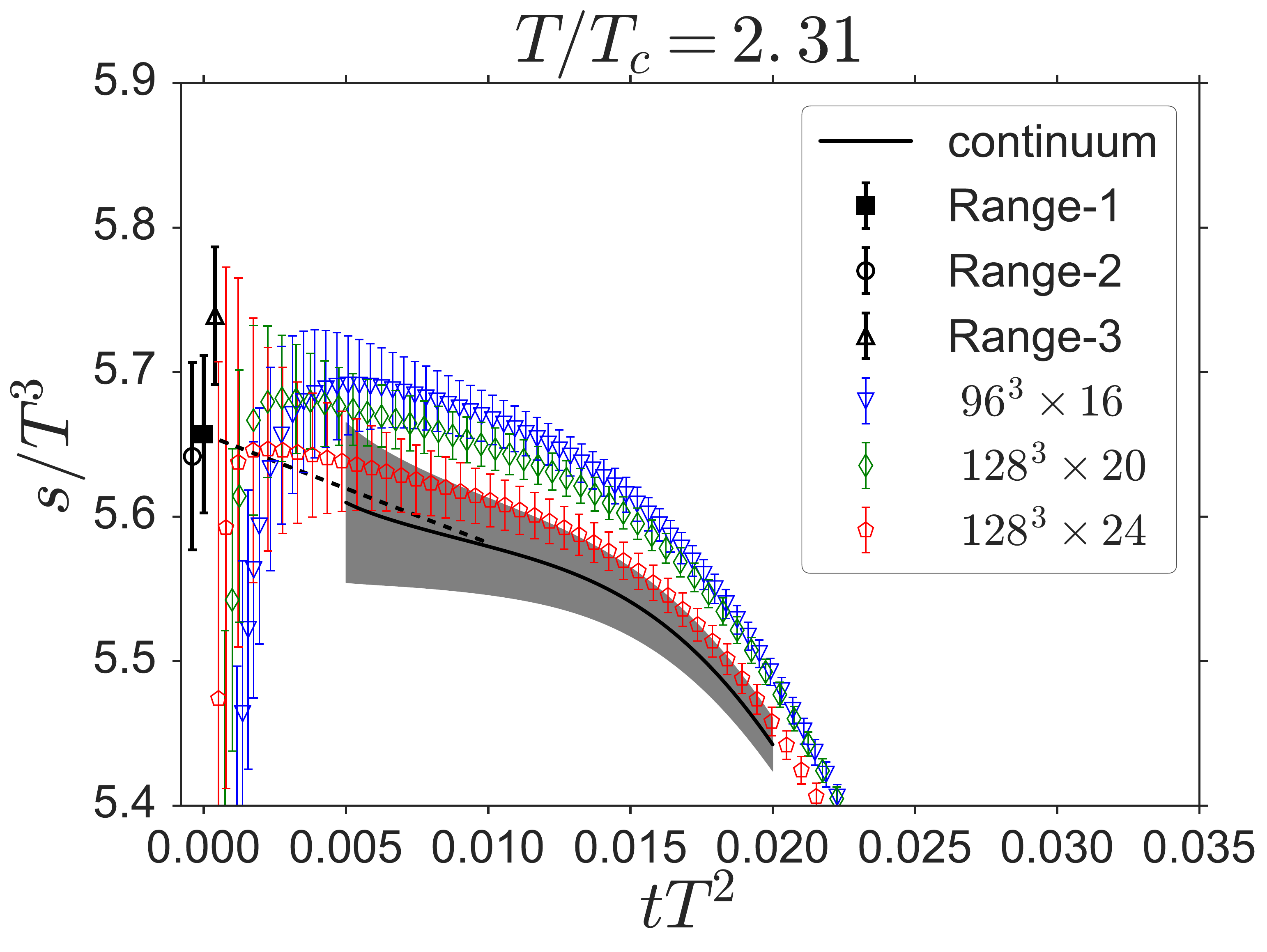}
  \includegraphics[width=0.235\textwidth,clip]{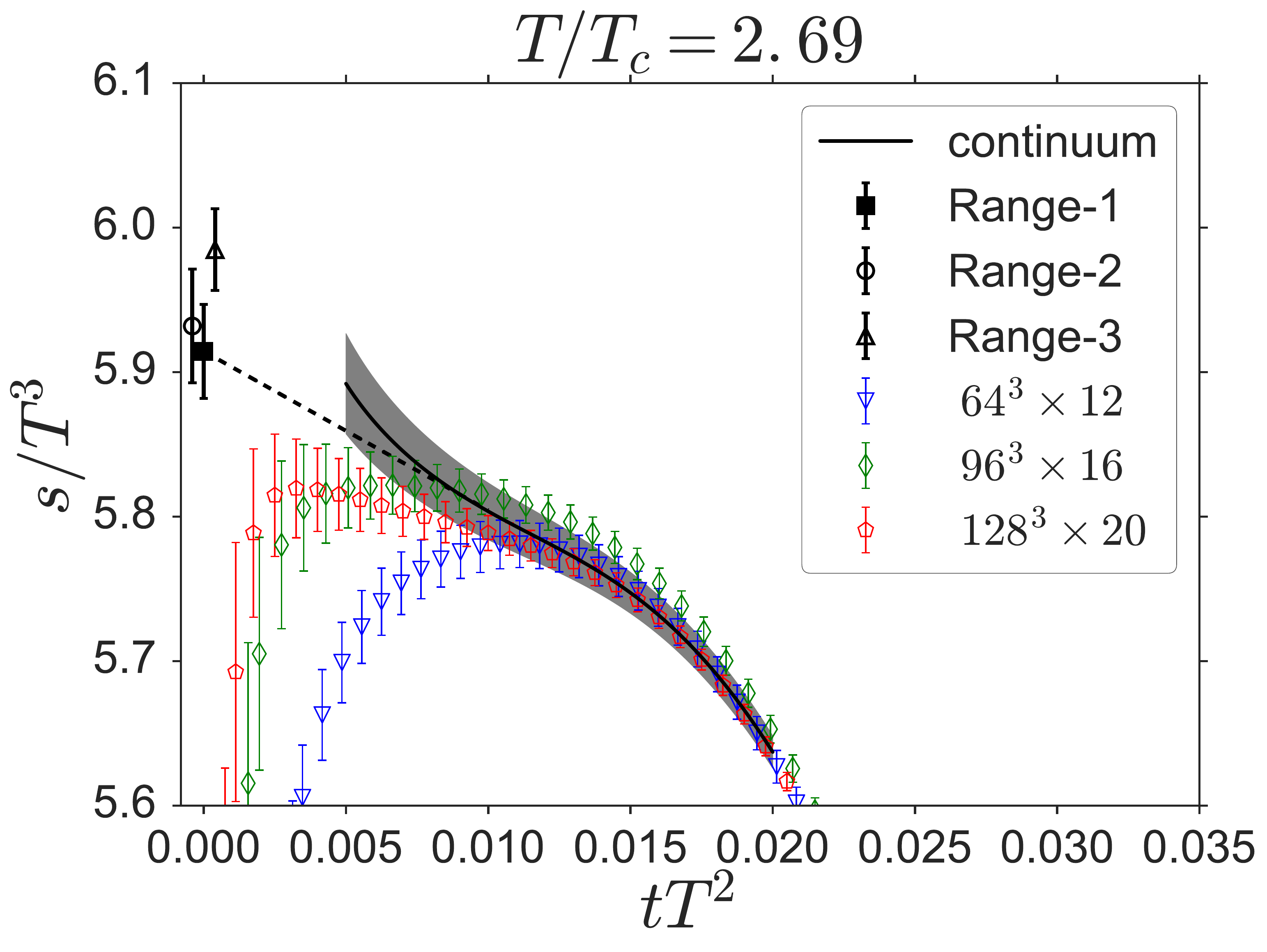}
\caption{
  Similar plots with the right panel of Fig.~\ref{fig:166cont} for different values of $T/T_c$.
}
\label{fig:e+p}
\end{figure}

\begin{figure*}
  \centering
  \includegraphics[width=0.45\textwidth,clip]{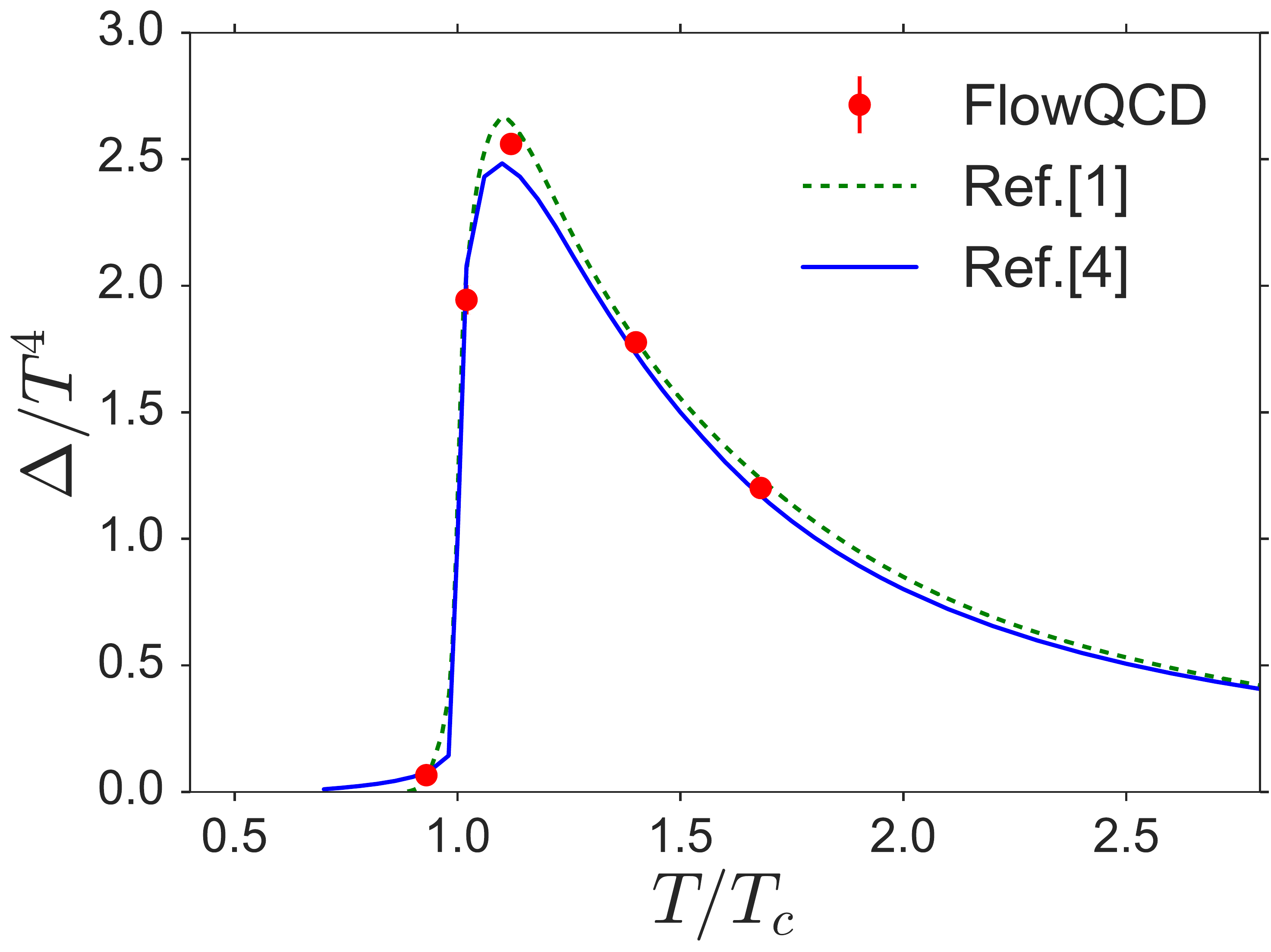}
  \includegraphics[width=0.45\textwidth,clip]{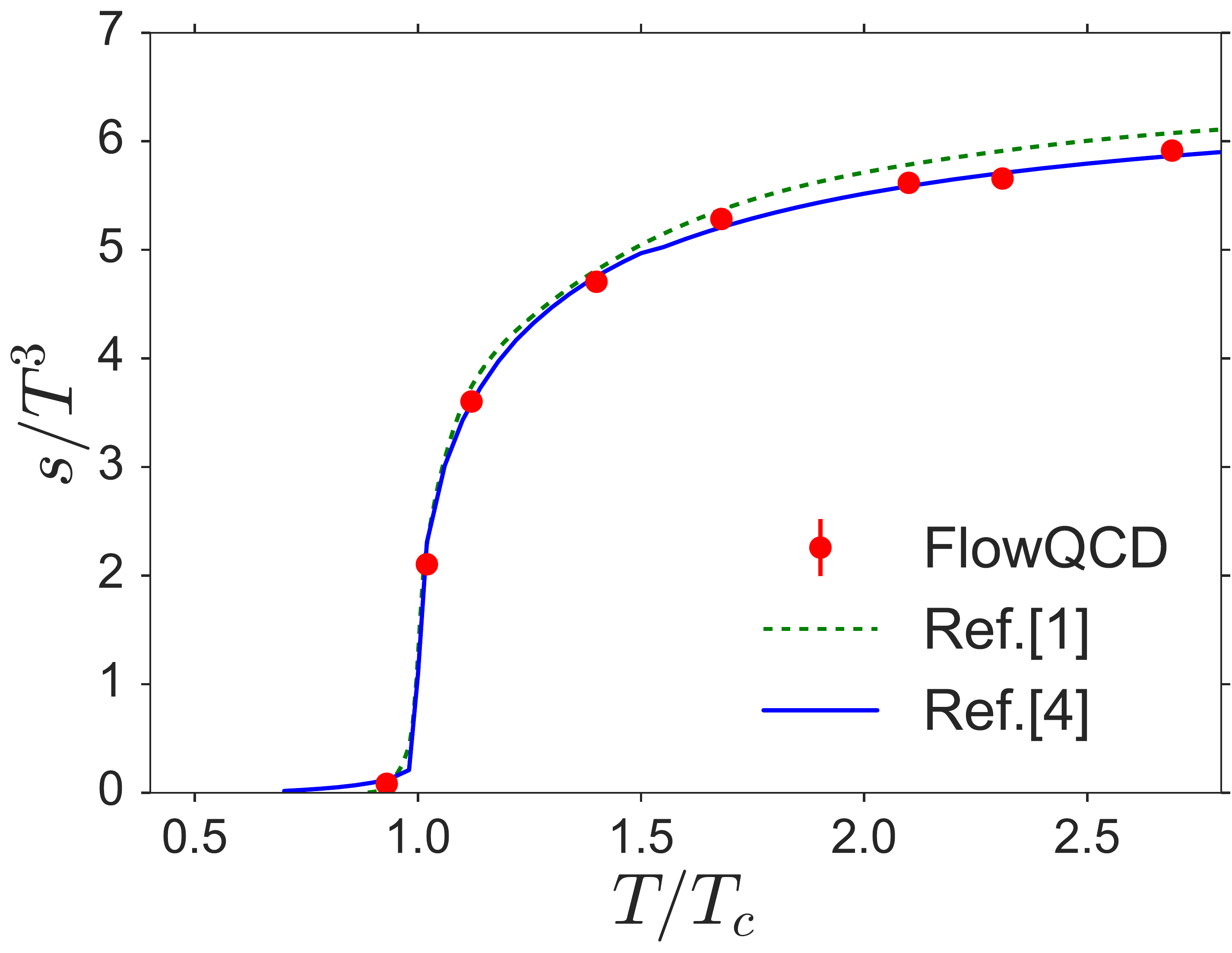}
  \caption{
    Temperature dependences of $\Delta/T^4$ and $s/T^3$ (red circles)
    together with the previous studies based on the integral method 
    (solid and dashed lines)
    \cite{Boyd:1996bx,Borsanyi:2012ve}.
    The error bars of  the red circles are  smaller than the size of symbols.
  \label{fig:Tdep}
  }
\end{figure*}

Let us now describe the procedure for the double extrapolation 
$(t,a)\to(0,0)$.
As discussed in Sec.~\ref{sec:setup},
we first take the continuum limit with $t$ fixed.
This extrapolation is taken by fitting the results with 
three different values of $N_\tau$ with Eq.~(\ref{eq:a->0}).
 To obtain the values of
 $\langle {T}_{\mu\nu}(t,x)\rangle_{\rm lat}$ at the same $t$ for different $N_{\tau}$,
we apply the cubic spline interpolation to the data for each $N_{\tau}$.

In Figs.~\ref{fig:cont_limit_anomaly} and \ref{fig:cont_limit_entropy}, 
we show the $N_\tau$ dependences of 
$\Delta/T^4$ and $s/T^3$ at  $tT^2 = 0.005$, $0.01$, $0.015$ and~$0.02$
together with the result of continuum extrapolation with 
Eq.~(\ref{eq:a->0}).
In Fig.~\ref{fig:cont_limit_anomaly}, 
three results obtained by the different choices for the operator $E(t,x)$ 
are shown.
The value of $\chi^2/{\rm dof}$ is within the range 
$\chi^2/{\rm dof} \lesssim 2.0$
for $0.005 \le tT^2 \le 0.02$.
The error of the continuum extrapolation is estimated by the jackknife analysis.
For values of $tT^2$ smaller than $0.005$, 
the fitting becomes worse particularly for $s/T^3$.
Therefore, in the following, we will use the results only for 
$0.005 \le tT^2 \le 0.02$.
Figure~\ref{fig:cont_limit_anomaly} also shows that the 
continuum extrapolated results with different discretizations for 
$E(t,x)$ agree with each other.

In Fig.~\ref{fig:166cont}, we show the $t$ dependences of $\Delta/T^4$ and
$s/T^3$  after the continuum extrapolation 
by the black line with the error band
together with the data for finite lattice spacings,  $N_\tau=12$, $16$ and $20$.
We make linear  $t$  extrapolation by using the 
 continuum extrapolated data for $0.005 \le tT^2 \le 0.02$
  according to Eq.~(\ref{eq:t->0}).
We employ three fitting ranges, 
\begin{description}
 \item[Range-1] $ 0.01 \le tT^2 \le 0.015$, 
 \item[Range-2] $ 0.005 \le tT^2 \le 0.015$, 
 \item[Range-3] $ 0.01 \le tT^2 \le 0.02$.
\end{description}
  In Fig.~\ref{fig:166cont},
  the black solid bar at $t=0$ with a squared symbol denotes
  the result of the extrapolation with Range-1,
  while 
  the open circle and triangle symbols
  denote the results with Range-2 and Range-3, respectively.
$\chi^2/{\rm dof}$ in these fittings  is smaller than unity.
Then, we use the result of Range-1 as a central value, while those of
 Range-2 and Range-3 are used to estimate the systematic error associated 
with the fit range.\footnote{In our previous exploratory study of 
$\Delta/T^4$ and $s/T^3$ in Ref.~\cite{Asakawa:2013laa}, 
the  continuum limit has been taken, while 
the flow time was fixed to be $tT^2=0.02$.
There was no resolution to detect the slope $C_{\mu \nu}$ 
owing to limited statistics and coarse lattice.}

\begin{table}
  \centering
  \begin{tabular}{c|ll}
    \hline \hline
$T/T_c$ & $\Delta/T^4$ & $s/T^3$ \\
\hline
 0.93 &    $0.066(32)(^{+3}_{-2})(0)$ &       $0.082(33)(^{+ 3}_{- 6})( 0)$ \\
 1.02 &    $1.945(57)(^{+8}_{-7})(0)$ &       $2.104(63)(^{+16}_{- 2})( 8)$ \\ 
 1.12 &    $2.560(33)(^{+12}_{-8})(0)$ &      $3.603(46)(^{+39}_{- 0})(13)$ \\ 
 1.40 &    $1.777(24)(^{+14}_{-3})(0)$ &      $4.706(35)(^{+49}_{- 0})(17)$ \\ 
 1.68 &    $1.201(19)(^{+10}_{-0})(0)$ &      $5.285(35)(^{+44}_{- 0})(18)$ \\ 
 2.10 &  \multicolumn{1}{c}{---}       & $5.617(34)(^{+66}_{- 0})(18)$ \\ 
 2.31 &  \multicolumn{1}{c}{---}       & $5.657(55)(^{+82}_{-15})(18)$ \\ 
 2.69 &  \multicolumn{1}{c}{---}       & $5.914(32)(^{+70}_{- 0})(18)$ \\ 
\hline
\hline
  \end{tabular}
  \caption{Summary of the equation of state  with statistical and systematic errors.
    The first error is the statistical one, while
    the second error shows the systematic error associated with 
    the choice of the fit range.
    The last error comes from $1\%$ uncertainties
  of $\Lambda_{\overline{\mathrm{MS}}}$ from possible topological freezing.
      $\Delta/T^4$  at $T/T_c = 2.10$, $2.31$ and $2.69$ are not available
      due to the lack of corresponding vacuum configurations.
    }
\label{tab:extrapol}
\end{table}

In order to estimate the systematic error from the 
uncertainly of   $a\Lambda_{\overline{\mathrm{MS}}}$ discussed in~Sec.~\ref{sec:setup}, 
we show the continuum extrapolated results 
 under $\pm 1\%$ change of  $a\Lambda_{\overline{\mathrm{MS}}}$
 in Fig.~\ref{fig:lambda_sys}.
As the figure shows, the systematic error for $\Delta/T^4$ ($s/T^3$) 
is  negligible (comparable) to the other statistical and systematic errors.

\subsection{Temperature dependence}

The analysis in the previous subsection for $T/T_c=1.68$ is 
repeated for all $T/T_c$ listed in~Table~\ref{table:param1}.
We show the results
of these analyses with different values of $T/T_c$ in Fig.~\ref{fig:e-3p} 
for $\Delta/T^4$ and 
in  Fig.~\ref{fig:e+p} for $s/T^3$.
The values of $\chi^2/{\rm dof}$ are within a reasonable range
$\chi^2/\mathrm{dof} \lesssim 2$ for all fits with 
an exception for $s/T^3$ at $T/T_c = 1.40$.
As these figures show, the double extrapolation works rather stably 
for all $T/T_c$.

The numerical results after double extrapolation
 are summarized in Table~\ref{tab:extrapol}.
 The table shows that $\Delta/T^4$ and $s/T^3$ are determined 
 within $3\%$ precision including all systematic errors
 except for those at $T/T_c=0.93$.
Note that we do not have  $\Delta/T^4$ for the highest three temperatures
owing to  the lack of vacuum simulations needed to make vacuum subtraction
(see Table~\ref{table:param1}).

 Finally, we depict the $T/T_c$ dependences of our $\Delta/T^4$ and $s/T^3$
in Fig.~\ref{fig:Tdep} together with the previous data obtained by the integral method
in Refs.~\cite{Boyd:1996bx,Borsanyi:2012ve}.
By taking into the estimated errors of the previous results, three results are consistent
 with each other.\footnote{We note that $s/T^3$ recently studied in the shifted 
boundary method \cite{Giusti:2012yj,Giusti:2015got} also seems to agree.}

\section{Summary}

We performed 
measurements of thermodynamic
quantities of the SU(3) Yang--Mills theory
from the direct analysis of 
the expectation value of energy-momentum tensor (EMT), 
Eq.~(\ref{eq:T^R}), constructed by
the Yang--Mills gradient flow with a flow time $t$.
The numerical simulations with the Wilson plaquette gauge action
 have been performed at finite temperature 
with the lattice spacing $a= 0.013$--$0.061\,\mathrm{fm}$
and the aspect ratio, $5.33 \le N_s/N_{\tau} \le 8$.

Using the lattice data, the double extrapolation ($t\to0$ after $a\to0$)
has been performed to obtain 
the interaction measure $\Delta(T)$ and the entropy density $s(T)$  
with a few percent precision including statistical and  systematic errors.
The results agree quite well 
with the previous high-precision data using the integral method.

The present approach with EMT  provides a new tool not only to 
calculate QCD equation of state accurately  but also 
to study correlation functions and transport coefficients of the 
quark-gluon plasma with firm theoretical basis.
The first step along these directions will be reported in the
forthcoming paper \cite{FlowQCD_New}.

\acknowledgments

The authors thank E.~Itou for discussions 
in the early stage of this study.
Numerical simulation for this study was carried out on 
IBM System Blue Gene Solution at
KEK under its Large-Scale Simulation Program 
(Nos.~13/14-20, 14/15-08, 15/16-15). 
This work is supported in part by JSPS KAKENHI Grant Numbers 
24340054, 25287046, 25287066, 25800148, 26400272, 16H03982 and by
RIKEN iTHES Project.

\appendix

\section{Lattice spacing and $\Lambda$ parameter}
\label{sec:scale}

In this appendix, we summarize our analysis of 
the lattice spacing  and $\Lambda_{\overline{\mathrm{MS}}}$.
The numerical data used are those given
in~Ref.~\cite{Asakawa:2015vta}.
Possible error originating from the topological freezing is also mentioned.

\subsection{Reference scale and lattice spacing}
\label{sec:scale1}

Numerical simulations  of the SU(3) Yang--Mills
theory with the Wilson plaquette action were performed 
 on $N_s^4 = 64^4$--$128^4$ lattices under the periodic boundary condition.
The values of~$\beta=6/g_0^2$, $N_s$ and the number of 
configurations $N_{\mathrm{conf}}$ are summarized in the three left columns in~Table~\ref{tab:scale}.

\begin{table}
  \centering
  \begin{tabular}{cccccc}
    \hline \hline
    $\beta$ & $N_\mathrm{s}$ & $N_\mathrm{conf}$ & $w_0/a$ & $a$ [fm] & $N_s a$ [fm] \\
    \hline
  6.3 & 64 &  30 &  2.877(5)      &  0.058(4) &  3.72(22)  \\ 
  6.4 & 64 & 100 &  3.317(4)      &  0.050(3) &  3.22(19)  \\ 
  6.5 & 64 &  49 &  3.797(8)      &  0.044(3) &  2.81(17)  \\ 
  6.6 & 64 & 100 &  4.356(9)      &  0.038(2) &  2.45(15)  \\ 
  6.7 & 64 &  30 &  4.980(23)     &  0.034(2) &  2.15(13)  \\ 
  6.8 & 64 & 100 &  5.652(17)     &  0.030(2) &  1.89(11)  \\ 
  7.0 & 96 &  60 &  7.297(18)     &  0.023(1) &  2.20(13)  \\ 
  7.2 & 96 &  53 &  9.348(66)     &  0.018(1) &  1.71(10)  \\ 
  7.4 & 128 & 40 & 12.084(61)     &  0.014(1) &  1.77(11)  \\ 
    \hline \hline
  \end{tabular}
  \caption{
    Simulation parameters for scale setting, $\beta=6/g_0^2$, 
    the lattice size $N_s$, and the number of configurations~$N_{\mathrm{conf}}$, 
    as well as the numerical results of $w_0/a$.
    The lattice spacing $a$ and the physical length $N_s a$
    in physical unit determined from $w_0 = 0.1670(10)\,\mathrm{fm}$  \cite{Sommer:2014mea} 
    are also shown.
  }
  \label{tab:scale}
\end{table}

We adopt the reference scale  $w_0$ defined by \cite{Borsanyi:2012zs}
\begin{equation}
  \left. t\frac{d}{dt}t^2\langle E(t)\rangle\right|_{t=w_{0}^2}= 0.3,
  \label{eq:w_0}
\end{equation}
with the operator $E(t)$ constructed by  the clover-type representation 
of the flowed field~$G_{\mu\nu}^a$  at  time $t$.
We use the Wilson gauge action~$S_{\mathrm{YM}}$ for the flow 
equation in~Eq.~(\ref{eq:GF}), and 
each measurement is separated by $1000$ Sweeps.
The values of $w_0/a$ with statistical errors are 
summarized in the fourth column of Table~\ref{tab:scale}.
The lattice spacings in physical unit estimated by 
$w_0 = 0.1670(10)$~fm \cite{Sommer:2014mea} are also given 
in the table together with the physical lattice volume $L=N_sa$.
Extra error due to topological freezing  is estimated to be about 
$1\%$ level as discussed later.\footnote{
See Refs.~\cite{Luscher:2011kk,McGlynn:2014bxa,Endres:2015yca} for simulation strategies which are supposed
to avoid the topological freezing.}

\begin{figure}
  \centering
  \includegraphics[width=0.45\textwidth,clip]{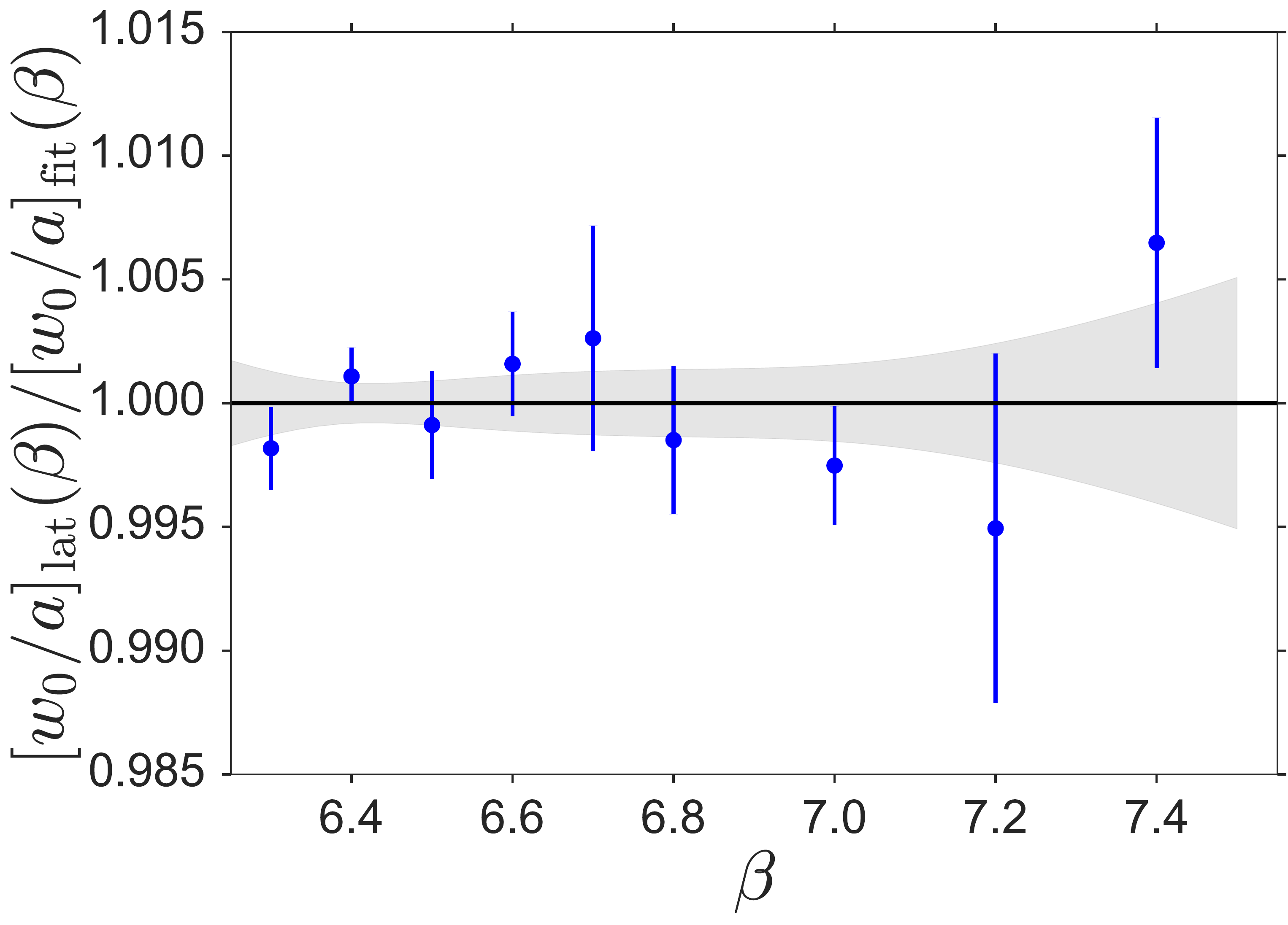}
  \caption{\label{fig:fit_scale_function} Result of the three parameter 
    fit of $w_0/a$ in Eq.~(\ref{eq:fitfinal}).
  The shaded band shows the uncertainties from the fit parameters.}
\end{figure}

For the parametrization of $w_0/a$ as a function of $\beta$,
we introduce the fitting function
motivated by the one-loop perturbation theory.
It provides a reasonable result ($\chi^2/{\mathrm{dof}}=1.104$) 
for $9$ data points 
in~$6.3 \le \beta \le 7.4$ without overfitting as shown in Fig.~\ref{fig:fit_scale_function}:
\begin{align}
   \frac{w_0}{a}
   =&\exp\left(\frac{4\pi^2}{33}\beta
     -9.1268
     +\frac{41.806}{\beta}
     -\frac{158.26}{\beta^2}
   \right)
\nonumber \\
&   [1\pm0.004(\mathrm{stat.})].
\label{eq:fitfinal}
\end{align}
    The difference from other fitting  Ans\"atze (such as polynomial functions
    as shown in Appendix A in Ref.~\cite{Asakawa:2015vta})
    is found to be less than $1\%$.
    The $0.4\%$ error in Eq.~(\ref{eq:fitfinal}) originates
     from  the statistical errors of $w_0/a$ except for the topological freezing.
   
For the analysis of $w_0/a$ in Table~\ref{tab:scale},
we have used $30$--$100$ configurations separated by $1000$ Sweeps.
In order to estimate the effect of the topological freezing 
on these simulations,  we have performed an independent measurement
at $\beta = 6.88$ on $N_s^4 = 64^4$ lattice
 by accumulating $N_\mathrm{conf} = 1290$
configurations with each measurement separated by $2000$ Sweeps.
This lattice setup corresponds to 
the physical size, $64 \times 0.027 \, \mathrm{fm} \simeq 1.7 \, \mathrm{fm}$, which is
comparable to the smallest lattice volume in Table~\ref{tab:scale}.
Since observables depend more on the topological sector
 for smaller spatial volume   \cite{Aoki:2007ka},
this analysis would serve as the most severe test for the topological freezing of
 the data sets in Table~\ref{tab:scale}.

\begin{figure}
  \centering
  \includegraphics[width=0.45\textwidth,clip]{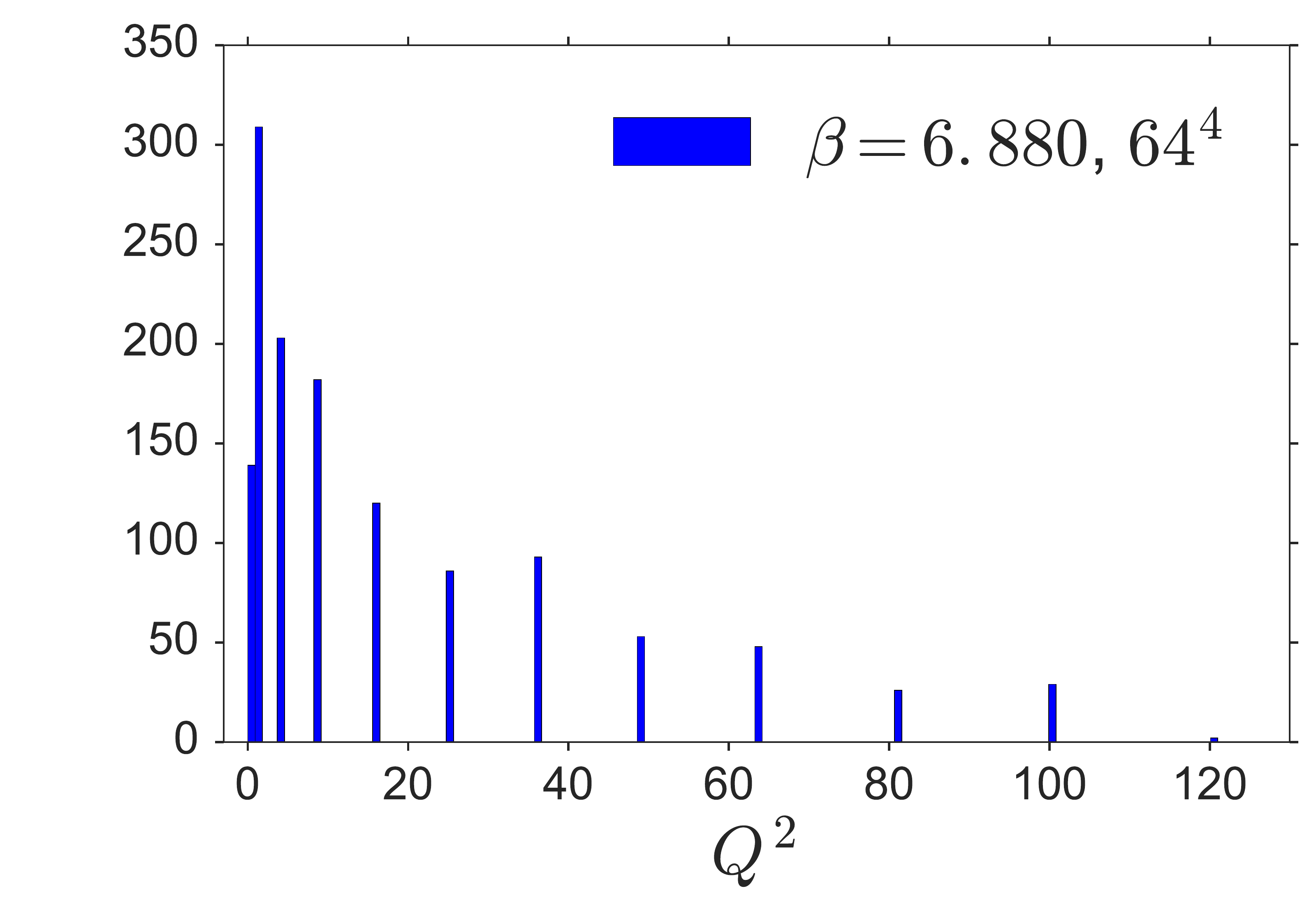}
  \includegraphics[width=0.45\textwidth,clip]{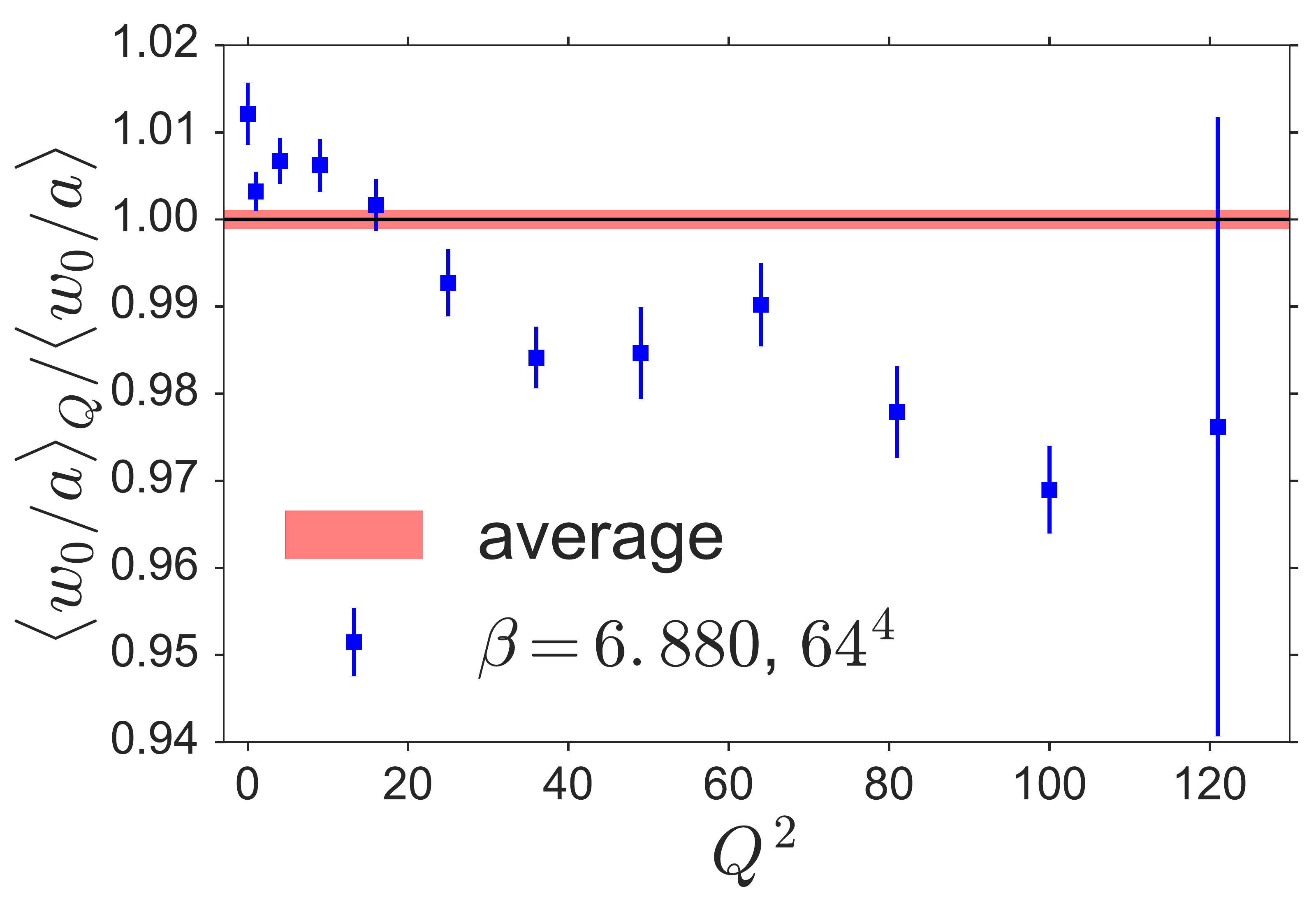}
  \caption{(a) Histogram of $Q^2$ for $\beta=6.88$ and $N_s^4 = 64^4$.
      (b)   The values of $w_0/a$ with fixed $Q^2$
       normalized by the expectation values using all configurations.
  }
\label{fig:w0_fixed_Q}
\end{figure}

The topological charge is defined by
$   Q \equiv -\frac{1}{32\pi^2}\epsilon_{\mu\nu\rho\sigma}
   \int_Vd^4x\, \mathrm{tr}  \left[G_{\mu\nu}(x)G_{\rho\sigma}(x)\right]$.
We take the value of $Q$ at $t = t_0$ defined by $t^2\left\langle E(t) \right\rangle|_{t=t_0}= 0.3$ \cite{Luscher:2010iy}. 
From this measurement of $Q$,
we find that the autocorrelation length is about $100 \times 2000$ Sweeps which
is $2$--$6$ times larger than the maximum number of Sweeps used to obtain $w_0/a$
in Table~\ref{tab:scale}.  Therefore, there is indeed a danger of the topological
freezing.
Shown in Fig.~\ref{fig:w0_fixed_Q}(a) is a histogram of $Q^2$ obtained 
 in the simulation.  The resultant fluctuation of $Q$ reads,
 $\langle Q^2 \rangle = 12.2 \pm 3.2$.
  Corresponding  topological susceptibility is estimated as
  $\chi a^4 \equiv \left\langle Q^2 \right\rangle/V =  (7.3 \pm 1.9) \times {10}^{-7}$,
with the error by the jackknife analysis with binsize 100.
By using Eq.~(\ref{eq:fitfinal}) 
and the reference values $w_0 = 0.1670(10)$ fm and $r_0 = 0.49$ fm \cite{Sommer:2014mea},
we find  $\chi r_0^4 = 0.084(22)$ which is in $1.5\sigma$ level of agreement with the accurate 
determination, $\chi r_0^4 = 0.0544(18)$ \cite{Ce:2015qha}.

In Fig.~\ref{fig:w0_fixed_Q}(b), we plot $w_0/a$ at fixed topology, 
$\langle w_0/a \rangle_Q$, normalized by the expectation value 
$\langle w_0/a\rangle$ without fixing $Q$.
The red band corresponds to the error of $\langle w_0/a\rangle$
with total configurations.
By combining the typical value expected from the topological 
 susceptibility  ($|Q|<\sqrt{\langle Q^2 \rangle} < 4$) 
 and the results of Fig.~\ref{fig:w0_fixed_Q},  
 we estimate the effect of the topological freezing is about $1\%$ level.
 
\subsection{Determination of $\Lambda_{\overline{\mathrm{MS}}}$}
\label{sec:Lambda}

For the value of $w_0\Lambda_{\overline{\mathrm{MS}}}$,
 we adopt a procedure similar to the one in~Ref.~\cite{Gockeler:2005rv}
for the determination of  $r_0\Lambda_{\overline{\mathrm{MS}}}$.
The dimensionless parameter $a\Lambda_{\overline{\mathrm{MS}}}$ can be 
obtained by matching the tadpole improved lattice perturbation theory. 
The boosted coupling constant~$g_\square$ is defined by
\begin{equation}
   g_\square^2\equiv g_0^2(a)/u_0^4,
\end{equation}
where $u_0^4\equiv P=\langle\mathrm{Tr}\ U_\square\rangle/3$.

As for the choice of the renormalization scale and the running coupling
constant, we take the following two methods:
\begin{itemize}
\item Method~I
\begin{equation}
   a\Lambda_{\overline{\mathrm{MS}}}
   =a\mu_\ast F^{\overline{\mathrm{MS}}}(g_{\overline{\mathrm{MS}}}(\mu_\ast))
\end{equation}
at the scale 
\begin{equation}
   a\mu_\ast=\exp\left(\frac{t_1^\square}{2b_0}\right),
\end{equation}
and 
\begin{equation}
   \frac{1}{g_{\overline{\mathrm{MS}}}^2(\mu_\ast)}
   =\frac{1}{g_\square^2(a)}
   +\left(\frac{b_1}{b_0}t_1^\square-t_2^\square\right)g_\square^2(a)
   +O(g_\square^4).
\end{equation}
\item Method~II
\begin{equation}
   a\Lambda_{\overline{\mathrm{MS}}}
   =a\Lambda_\square\exp\left(\frac{t_1^\square}{2b_0}\right),
\end{equation}
with 
\begin{equation}
   a\Lambda_\square=F^\square(g_\square(a)).
\end{equation}
This scheme corresponds to choosing a scale at
\begin{equation}
   a\mu_=
   =\exp\left(\frac{t_1^\square}{2b_0}\right)
   \frac{F^\square(g_\square(a))}{F^{\overline{\mathrm{MS}}}(g_\square(a)}
\end{equation}
in Method~I.
\end{itemize}
At the $3$-loop order, $F^S$ ($S=\square$, $\overline{\mathrm{MS}}$) is
expressed as
\begin{align}
   \frac{\Lambda^S}{M}
   &\equiv F^S(g_S(M))
   =\exp\left(-\frac{1}{2b_0g_S^2}\right)(b_0 g_S^2)^{-\frac{b_1}{2b_0}}
\notag\\
   &\qquad{}\times
   \left(1+\frac{b_1+\sqrt{b_1^2-4b_0b_s^S}}{2b_0}g_S^2\right)^{-p_A^S}
\notag\\
   &\qquad{}\times
   \left(1+\frac{b_1+\sqrt{b_1^2+4b_0b_s^S}}{2b_0}g_S^2\right)^{-p_B^S},
\end{align}
where
\begin{align}
   p_A^S
   =-\frac{b_1}{4b_0^2}-\frac{b_1^2-2b_0b_2^S}{4b_0^2\sqrt{b_1^2-4b_0b_2^S}},
\\
   p_B^S
   =-\frac{b_1}{4b_0^2}+\frac{b_1^2-2b_0 b_2^S}{4b_0^2\sqrt{b_1^2-4b_0b_2^S}}.
\end{align}
In the $[1,1]$ Pad\'e approximation, it leads to
\begin{equation}
   F_{[1,1]}^S(g_S(M))
   =\exp\left(-\frac{1}{2b_0 g_S^2}\right)
   \left[\frac{b_0g_S^2}{1+\left(\frac{b_1}{b_0}-\frac{b_2^S}{b_1}\right)g_S^2}
   \right]^{-\frac{b_1}{2b_0}}.
\end{equation}
In SU(3) Yang--Mills theory, the coefficients are given by
\begin{align}
&   b_0=\frac{11}{(4\pi)^2},\qquad b_1=\frac{102}{(4\pi)^4},\qquad
   b_2^{\overline{\mathrm{MS}}}=\frac{1}{(4\pi)^6}\frac{2857}{2}, 
\nonumber \\
&   b_2^\square=b_2^{\overline{\mathrm{MS}}}+b_1t_1^\square-b_0t_2^\square,
\end{align}
with
\begin{equation}
   t_1^\square=0.1348680,\qquad t_2^\square=0.0217565.
\end{equation}

\begin{figure}
  \centering
  \includegraphics[width=0.45\textwidth,clip]{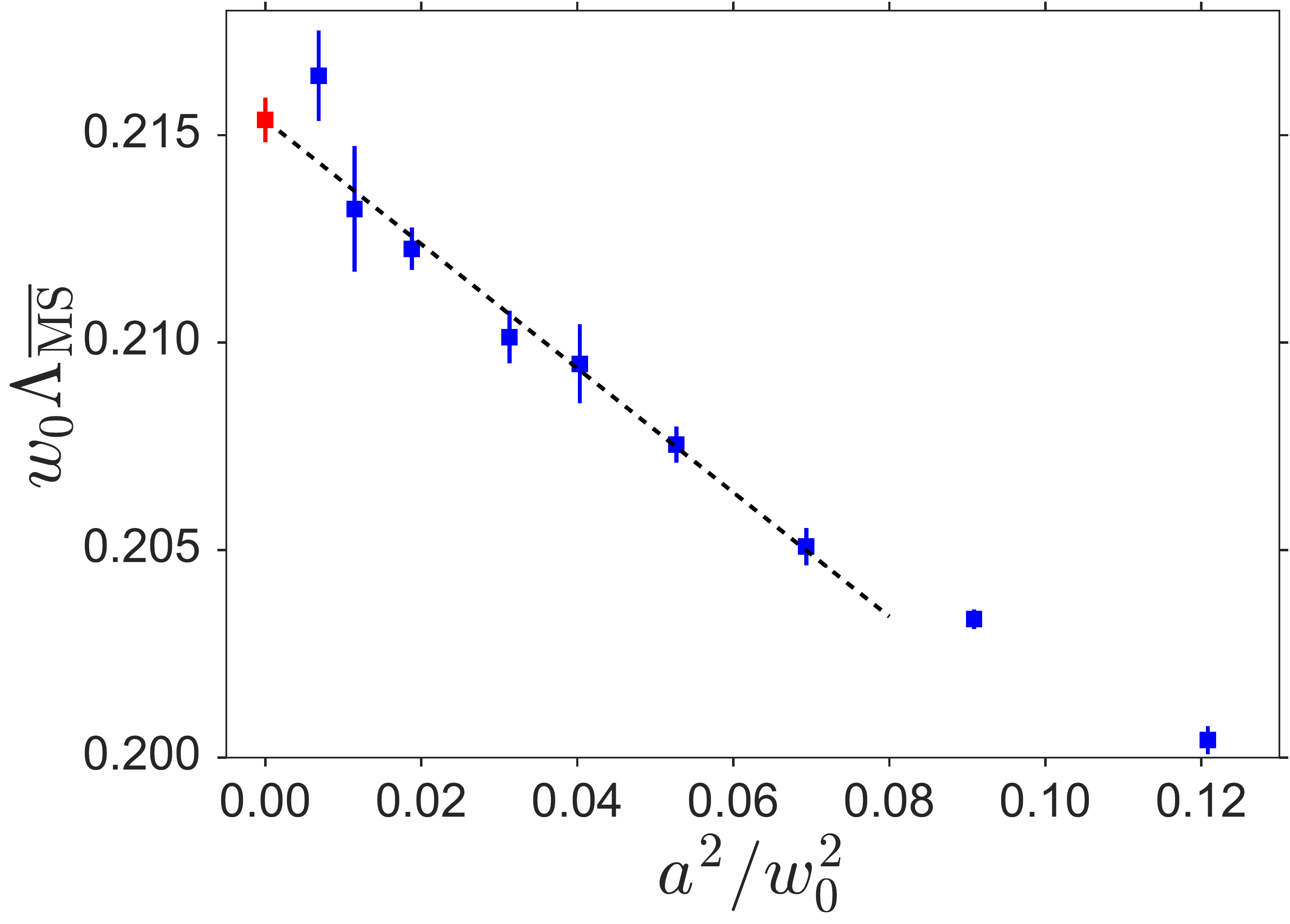}
  \caption{Values of $w_0 \Lambda_{\overline{\mathrm{MS}}}$ by the Method II with Pad\'e improvement
    as a function of lattice spacing $a^2$. 
    The continuum limit is shown at $a^2/w_0^2 = 0$.
    The finest lattice data at $\beta = 7.4$ deviates from fitting line.
    We note that the continuum extrapolation
    is consistent within the statistical error without using the result at $\beta = 7.5$.
    \label{fig:lambdaMSbarIIPade}
  }
\end{figure}

The expectation values of the plaquette,
$w_0$ and $w_0 \Lambda_{\overline{\mathrm{MS}}}$ with three schemes 
are summarized in Table~\ref{tab:lambdaMSbar}.
Following Ref.~\cite{Gockeler:2005rv},
we adopt Method~II with Pad\'e improvement to estimate 
the central value of~$w_0\Lambda_{\overline{\mathrm{MS}}}$. 
The values in the continuum limit are obtained by a linear fit 
as a function of~$a^2/w_0^2$ without using the coarse results 
at $\beta = 6.3$ and~$6.4$ (see~Fig.~\ref{fig:lambdaMSbarIIPade}). 
We used the results of the other methods to estimate the systematic error.
From this analysis we find
\begin{equation}
   w_0\Lambda_{\overline{\mathrm{MS}}}=0.2154(5)(11).
\label{eq:wL}
\end{equation}
Note that  the topological freezing discussed in Appendix~\ref{sec:scale1}
would introduce another 1\% error  to this number.

\begin{table*}
\centering
\begin{tabular}{rrlrlll}
    \hline \hline
    $\beta$ & $N_{\rm s}$ & \multicolumn{1}{c}{plaquette} & \multicolumn{1}{c}{$w_{0}/a$} & \multicolumn{3}{c}{$w_{0} \Lambda_{\overline{\mathrm{MS}}}$} \\ \cline{5-7}
      & & & & Method I & Method II & Method II Pad\'e \\ 
    \hline
6.3 & 64 &  0.622 420 85(30)  &  2.877(5)  & \textit{0.2017(3)}&\textit{0.2021(3)}&\textit{0.2004(3)}\\
6.4 & 64 &  0.630 632 88(13)  &  3.317(4)  & \textit{0.2046(2)}&\textit{0.2050(2)}&\textit{0.2033(2)}\\
6.5 & 64 &  0.638 361 33(35)  &  3.797(8)  & 0.2063(5)  &  0.2067(5)  & 0.2051(4)  \\
6.6 & 64 &  0.645 669 58(12)  &  4.356(9)  & 0.2087(4)  &  0.2091(4)  & 0.2075(4)  \\
6.7 & 64 &  0.652 608 39(39)  &  4.980(23) & 0.2106(10) &  0.2109(10) & 0.2095(10) \\
6.8 & 64 &  0.659 215 11(11)  &  5.652(17) & 0.2112(6)  &  0.2115(6)  & 0.2101(6)  \\
7.0 & 96 &  0.671 556 729(89) &  7.297(18) & 0.2133(5)  &  0.2136(5)  & 0.2123(5)  \\
7.2 & 96 &  0.682 891 86(22)  &  9.348(66) & 0.2142(15) &  0.2144(15) & 0.2132(15) \\
7.4 & 128 & 0.693 365 795(68) & 12.084(61) & 0.2173(11) &  0.2176(11) & 0.2164(11) \\
\hline
$\infty$ & & 1 & $\infty$ &     0.2163(5)  &   0.2165(5)  &   0.2154(5)  \\
& & & ($\chi/$dof)        &      (0.927) &  (0.902)  & (0.991) \\
    \hline \hline
\end{tabular}
\caption{Simulation parameters $\beta$ and~$N_{\rm s}$. The plaquette value,
$w_0/a$ and~$w_0\Lambda_{\overline{\mathrm{MS}}}$ using Method I, II and~II
with Pad\'e approximation. The last row corresponds to the values at the
continuum limit obtained from linear extrapolation without using two coarse
lattice data at~$\beta=6.3$ and 6.4 (the italic numbers).}
\label{tab:lambdaMSbar}
\end{table*}

\end{document}